\documentclass[%
reprint,
superscriptaddress,
showkeys,
amsmath,amssymb,
pra,
longbibliography,
floatfix,
nobalancelastpage
]{revtex4-2}

\usepackage{defs}

\begin{document}
\title{General Time-Dependent Surface Flux for
  Multiple Ionization and Fragmentation of Molecules\\I: General Theory}

\author{Stefanos~Carlström\,\orcidlink{0000-0002-1230-4496}}%
\email{stefanos@mbi-berlin.de}
\affiliation{Max-Born-Institut, Max-Born-Straße 2A, 12489 Berlin, Germany}

\author{Serguei~Patchkovskii}%
\affiliation{Max-Born-Institut, Max-Born-Straße 2A, 12489 Berlin, Germany}

\date{\today}

\begin{abstract}
  We introduce a method for efficiently computing multiple
  photofragmentation of atoms and molecules into photoelectrons and
  nuclear fragments. The method is a generalization of the
  time-dependent surface flux originally introduced for single and
  double photoionization.
\end{abstract}


\maketitle

\section{Introduction}

The computation of photoelectron and kinetic energy release (KER)
spectra in photoionization and photodissociation of molecules can be a
daunting task, since the correlated fragments formally span all of
space. Even in a limited numeric expansion of the wavefunction,
finding the scattering states is costly, since diagonalization of the
Hamiltonian scales as \(\Ordo\{N^3\}\), where \(N\) is the number of
expansion coefficients. Therefore, since the earliest calculations of
photofragmentation spectra, various methods have been developed to
overcome this computational bottleneck.

Perhaps the simplest approach is to apply the \emph{fast Fourier
  transform} (FFT) to the final-time wavefunction (typically masking
out the central region, where the remaining bound part of the
wavefunction is), to obtain a momentum-space representation
\cite{Morales2011}. One drawback of this approach is necessity for
keeping the full wavefunction inside the computational box,
potentially leading to expensive calculations and large storage
requirements. Another drawback is the neglect of the binding
potential, leading to subtle energy shifts of photofragment peaks.

Another class of methods are the so-called \emph{window methods}
\cite{Schafer1990,Schafer1991CPC,Catoire2012,Silva2013,Catoire2014},
where a simple function of the resolvent \([\Hamiltonian-E]^{-1}\) is
applied to the final wavefunction:
\(P(E) = \braket{\chi(E)}{\chi(E)}\), where
\(\ket{\chi(E)} = a[f(\Hamiltonian-E)]^{-1}\ket{\Psi(t_f)}\). The main
benefit of this approach is that the exact scattering state with
energy \(E\) of the Hamiltonian \(\Hamiltonian\) is found via the
resolvent, essentially a variation of inverse iterations for finding
eigenstates. For most Hamiltonians this can be done with complexity
\(\Ordo\{N\}\). Again, the drawback is that the entire wavefunction
needs to be kept in the computational box.

It is possible to decrease the computational box to a much smaller
volume, and still compute accurate photofragmentation spectra, if one
transfers all outgoing flux that is removed from the box (via
absorbing boundary conditions) to an auxiliary representation of the
wavefunction. Early such \emph{grid transfer methods} employed a
finite matching region, where both the original and the auxiliary
representations overlapped \cite{Chelkowski1998}. An interesting
variation of this employs the complex-absorbing potential itself as
the matching region \cite{Selstoe2021,Selstoe2022,Dalen2025}. Via the
divergence theorem, it is possible to performing this matching at one
surface only, reducing the volume integral to a surface integral
instead. This lays the ground for the \emph{surface-flux} family of
techniques
\cite{Ermolaev1999,Ermolaev2000,Tao2012NJoP,Scrinzi2012NJoP,Morales2016-isurf,Orimo2019,Carlstroem2022tdcisI}. In
this paper, we introduce a systematic generalization of the
surface-flux technique to tackle photofragmentation into multiple
levels of fragmentation, \emph{simultaneously}. In doing so, we pave
the way for future studies on correlated multiple photoionization and
photodissociation of molecules.

The rest of this paper is organized as follows: in
section~\ref{sec:theory}, the rather extensive theory behind the
proposed method is discussed in some detail, with interspersed
examples illustrating the general method in specialized cases; in
section~\ref{sec:demonstration}, the method is applied to a numerical
example; finally, section~\ref{sec:discussion} concludes the paper
with a discussion.

\subsection{Conventions}
We describe the generalization of the t+iSURFF to multiple ionization
and molecular fragmentation, and term this fSURFF. To keep the
terminology light, we will use the word \enquote{fragment} to refer to
both photoelectrons and molecular (ionic) fragments.

We employ a variant of Einstein's summation convention, where all
quantities appearing on one side only of an equality sign is
implicitly contracted over. In Euclidean space \(\Real^N\), we use
ordinary vector calculus notation, \(\vec{r}\) is a vector,
\(\vec{k}\cdot\vec{r}=\eta_{ij}k^ir^j\equiv k_ir^i\) is the inner
product with the flat metric \(\eta_{ij}=\delta_{ij}\), \(\nabla\) and
\(\partial_x\) are derivative operators. On Riemannian manifolds, we
use common index notation, e.g.\ the partial derivative of a vector
field is written \(\tensor{V}{^a_{,b}}\equiv\partial_bV^a\).

We also use Hartree atomic units, wherein
\(\hbar = \elementarycharge = \electronmass = 4\pi\varepsilon_0 = 1\),
the charge of the electron is
\(\electroncharge = -\abs{\elementarycharge} = -1\), and the unit of
energy is \(\SI{1}{\hartree}\approx\SI{27.211}{\electronvolt}\). The
time-dependent Schrödinger equation is abbreviated TDSE.

\section{Theory}
\label{sec:theory}
\subsection{Partitioning of Space}
\label{sec:space-partitioning}
The basic idea behind the surface flux technique is the partitioning
of space into an interior and an exterior, similar to the \(R\)-matrix
technique \cite{Burke2011}. In the one-dimensional case first
described by \textcite{Tao2012NJoP}, this partitioning is simply
accomplished by a Heaviside function, shifted to the matching point
\(x_0\):
\begin{equation}
  \label{eqn:one-dimensional-partitioning}
  \braket{u}{v}_{\abs{x}>x_0} \defd
  \int\diff{x}
  \conj{u}(x)
  \heaviside(\abs{x}-x_0)
  v(x),
\end{equation}
where
\begin{equation}
  \label{eqn:heaviside}
  \heaviside(x-x_0) =
  \begin{cases}
    1, & x \ge x_0, \\
    0, & x < x_0.
  \end{cases}
\end{equation}
The generalization to higher dimensions is obvious in the spherically
symmetric case; the matching point becomes a matching hypersphere of
radius \(r_0\) centred at the origin
\cite{Morales2016-isurf,Carlstroem2022tdcisII}. To generalize to other
symmetries, or cases where the is no symmetry present, we will
introduce the partitioning of space into a general domain
\(\domain\subset\Real^M\) and its complement
\(\complementdomain\defd\Real^M\setminus\domain\), where
\(M\defd nN\), \(n\) is the number of dimensions of the universe
(typically \(n=3\)), and \(N\) is the total number of fragments. We
only require that \(\domain\) is connected; the generalization to a
finite set of disjoint domains is trivial. In practice, the choice of
\(\domain\) is dictated by the computational basis used; in
(hyper)spherical coordinates a (hyper)sphere is the obvious choice, in
Cartesian coordinates an axis-aligned box, \&c. Typically, we wish to
use a set of primitive domains for the various fragments, and we
construct the collective domain as the tensor product of the primitive
domains:
\begin{equation}
  \domain = \bigotimes_{i=1}^N \primitivedomain(i),
\end{equation}
where \(\primitivedomain(i)\subset\Real^n\) denotes the primitive domain of the
\(i\)th fragment. An exception could be when treating double
ionization of \ch{He} in hyperspherical coordinates
\cite{Herrick1975,Herrick1975a,Herrick1978,Herrick1980,Herrick1980b,
  Kellman1980,Rettrup1980,Lin1981,Lin1982,Lin1982b,Lin1984,Lin1995,
  Rost1997,Argenti2007,Argenti2007a}, where we may wish to explicitly
define \(\domain\) as a subset of \(\Real^6\). The derivation below is
equally valid for \(\domain\) and \(\primitivedomain(i)\).


\begin{figure}[htb]
  \centering
  \includegraphics[trim=30cm 30cm 30cm 30cm]{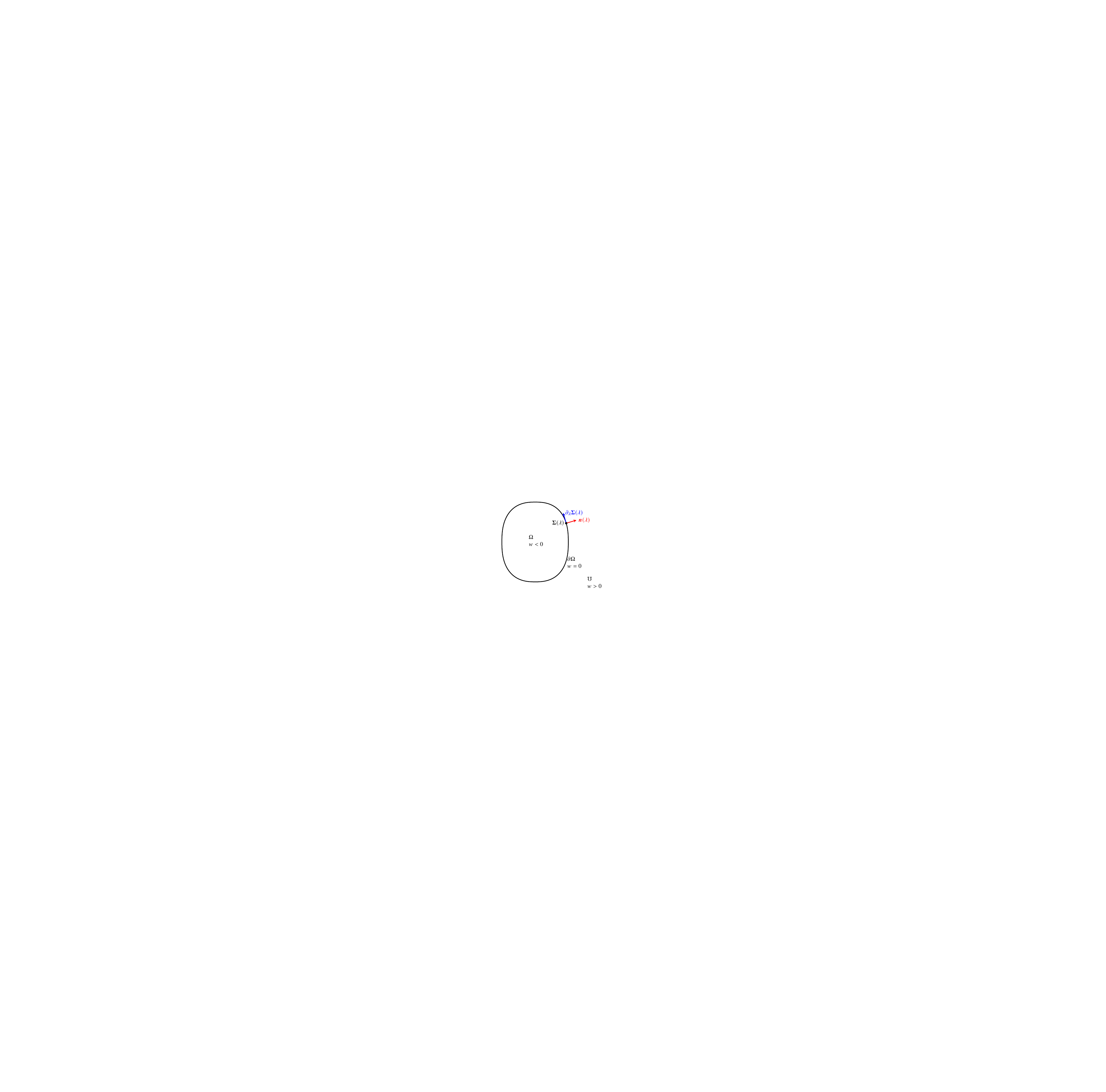}
  \caption{Illustration of the space partitioning and the local
    coordinate system.}
  \label{fig:local-coordinate-system}
\end{figure}
\(\vec{r}\) is any point in \(\Real^M\). Similarly,
\(\vec{r}_0 \in \domainboundary\) is any point on the surface
\(\domainboundary\) of the domain \(\domain\). For any point
\(\lambda\in\Real^{M-1}\), there is a mapping (coordinate chart)
\begin{equation}
  \label{eqn:surface-mapping}
  \vec{\Sigma} : \Real^{M-1} \to \domainboundary
\end{equation}
from that point to a corresponding \(\vec{r}_0\). This mapping is not
necessarily an isomorphism, and may have singular points. The
prototypical example is spherical coordinates, which map the two
angles \((\theta,\phi)\) to the unit sphere; this mapping is singular at the
poles \(\theta=0,\pi\). This is however not an issue for us, since we assume
the surface \(\domainboundary\) is smooth enough to admit setting up a
local, rectilinear coordinate system around the point
\(\vec{r}_0 = \vec{\Sigma}(\lambda)\). To this end, we find the tangent plane
from the variation of the mapping
\(\partial_\lambda\vec{\Sigma}(\lambda)\) around the point
\(\lambda\). Additionally, we need a normal vector orthogonal to this plane,
which we denote \(\vec{n}(\lambda)\) with \(\abs{\vec{n}(\lambda)}=1\), such that
\begin{equation}
  \label{eqn:projection-on-surface-normal}
  \begin{aligned}
    w
    &\defd
      (\vec{r}-\vec{r}_0)\cdot\vec{n},
    &
      \begin{cases}
        w > 0, & \vec{r}\textrm{ exterior to }\domain, \\
        w = 0, & \vec{r}\textrm{ on }\domainboundary, \\
        w < 0, & \vec{r}\textrm{ interior to }\domain.
      \end{cases}
  \end{aligned}
\end{equation}

With the introduction of one more parameter \(\zeta\), any point in
\(\Real^N\) may be written as
\begin{equation}
  \label{eqn:any-point-parametrization}
  \vec{r} = \vec{\Sigma}(\lambda) + \zeta\vec{n}(\lambda),
\end{equation}
i.e.\ as a point on the ray starting at the surface
\(\domainboundary\) and directed along its normal at that point; see
figure~\ref{fig:local-coordinate-system}. For convex domains
\(\Omega\), this mapping is unique in
\(\domainboundary\cup\complementdomain\); for non-convex domains we
may find points in \(\Real^M\) where two or more rays emanating from
different points \(\{\lambda\}\) on \(\domainboundary\)
\enquote{cross}, leading to multiple coverage. However, in an
\(\varepsilon\)-neighbourhood of \(\domainboundary\), the mapping is
unique, which is sufficient for the derivation below.

We introduce the bilinear form that maps onto the complement domain
\(\complementdomain\) as
\begin{equation}
  \label{eqn:complement-mapping-operator}
  \matrixel{u}{\complementmapping}{v}
  \defd
  \int_\complementdomain\diff{\vec{r}}
  \conj{u}(\vec{r})
  v(\vec{r}),
\end{equation}
where \(\vec{r}\) is given by
equation~\eqref{eqn:any-point-parametrization}. This is our
generalization of equation~\eqref{eqn:one-dimensional-partitioning} to
arbitrary domains \(\domain\), the mapping onto which is trivially
given by
\begin{equation}
  \label{eqn:mapping-operator}
  \mapping
  \defd
  \identity - \complementmapping
  \implies
  \matrixel{u}{\mapping}{v}
  \defd
  \int_\domain\diff{\vec{r}}
  \conj{u}(\vec{r})
  v(\vec{r}).
\end{equation}
Note that \eqref{eqn:complement-mapping-operator} is valid for
multi-component functions \(u,v\), such as those appearing when
solving the Pauli--Schrödinger equation.

\subsection{Probability in \(\complementdomain\)}
\label{sec:probability-in-complement-domain}
The probability of finding the system in \(\complementdomain\) is
given by
\begin{equation}
  \label{eqn:probability-in-complement-domain}
  \matrixel{\Psi(T)}{\complementmapping}{\Psi(T)} =
  \int_{-\infty}^T\diff{t}
  \partial_t
  \matrixel{\Psi(t)}{\complementmapping}{\Psi(t)},
\end{equation}
where we have employed the other important aspect of the surface flux
technique: by differentiating and integrating in time, we trade the
volume integral, which is costly to evaluate, for a time and surface
integral, as we will see below. In writing
\eqref{eqn:probability-in-complement-domain}, we have assumed the
initial condition
\(\matrixel{\Psi(-\infty)}{\complementmapping}{\Psi(-\infty)} = 0\),
and that \(\abs{\Psi}^2\) is \(L^2\) integrable.
\begin{widetext}
  \begin{equation}
    \label{eqn:complement-domain-probability}
    \begin{aligned}
      &\partial_t
        \matrixel{\Psi}{\complementmapping}{\Psi}
        \equiv
        -\int_\domain\diff{\vec{r}}
        \conj{\Psi}
        (\partial_t\Psi) +
        (\partial_t\conj{\Psi})
        \Psi =
        \im
        \int_\domain\diff{\vec{r}}
        \left[\left(-\frac{\tilde{\partial}_{\vec{r}}^2}{2} + V +
        \im\vec{G}\cdot\partial_{\vec{r}}\right)\Psi\right]
        \conj{\Psi} -
        \left[\left(-\frac{\tilde{\partial}_{\vec{r}}^2}{2} + V -
        \im\vec{G}\cdot\partial_{\vec{r}}\right)\conj{\Psi}\right]
        \Psi \\
      &=
        \im
        \int_\domain\diff{\vec{r}}
        \left[-\frac{\tilde{\partial}_{\vec{r}}^2}{2}\Psi +
        \im\vec{G}\cdot\partial_{\vec{r}}\Psi\right]
        \conj{\Psi} +
        \left[\frac{\tilde{\partial}_{\vec{r}}^2}{2}\conj{\Psi} +
        \im\vec{G}\cdot\partial_{\vec{r}}\conj{\Psi}\right]
        \Psi =
        \im
        \int_\domain\diff{\vec{r}}
        \partial_{\vec{r}}\cdot
        \left[
        \Psi\frac{\tilde{\partial}_{\vec{r}}}{2}\conj{\Psi} -
        \conj{\Psi}
        \frac{\tilde{\partial}_{\vec{r}}}{2}\Psi +
        \im\conj{\Psi}\Psi\vec{G}\right] \defd
        \int_\domain\diff{\vec{r}}
        \partial_{\vec{r}}\cdot\tilde{\vec{j}}.
    \end{aligned}
  \end{equation}
\end{widetext}
where we have suppressed the spatial arguments for brevity, and have
introduced the \emph{mass-weighted} gradient and Laplacian
\begin{equation}
  \label{eqn:mass-weighted-differentials}
  \begin{aligned}
    \tilde{\partial}_{\vec{r}}
    &\defd
      \frac{\partial_{\vec{r}_n}}{m_n},
    &
    \tilde{\partial}_{\vec{r}}^2
    &\defd
      \frac{\partial_{\vec{r}_n}^2}{m_n}
      \equiv
      \partial_{\vec{r}_n}
      \cdot
      \tilde{\partial}_{\vec{r}},
  \end{aligned}
\end{equation}
where \(m_n\) is the mass of the particle associated with the \(n\)th
coordinate. In we employ the dipole approximation, \(\vec{G}\)
vanishes in the length gauge, and in all other cases it is given by
\begin{equation}
  -\frac{q_n}{m_n}\vec{A}(\vec{r},t)\cdot\vecop{p} =
  \im\frac{q_n}{m_n}\vec{A}(\vec{r},t)\cdot\partial_{\vec{r}} \defd
  \im\vec{G}\cdot\partial_{\vec{r}},
\end{equation}
where similarly, \(q_n\) is the charge of the particle associated with
the \(n\)th coordinate (the \(A^2\) term is here taken as a term
inside \(V\)). With the aid of the vector calculus identity
\(\nabla\cdot(\phi\vec{F}) = (\nabla\phi)\cdot\vec{F} +
\phi(\nabla\cdot\vec{F})\), we find
\begin{equation}
  \begin{aligned}
    &
    \partial_{\vec{r}}\cdot(\conj{\Psi}\Psi\vec{G}) =
      (\partial_{\vec{r}} \conj{\Psi}\Psi)\cdot\vec{G} +
      \conj{\Psi}\Psi(\partial_{\vec{r}}\cdot\vec{G}) \\
    &=
      (\conj{\Psi} \vec{G}\cdot\partial_{\vec{r}} \Psi +
      \Psi \vec{G}\cdot\partial_{\vec{r}} \Psi) +
      \conj{\Psi}\Psi(\partial_{\vec{r}}\cdot\vec{G}) \\
    &\conj{\Psi} \vec{G}\cdot\partial_{\vec{r}} \Psi +
      \Psi \vec{G}\cdot\partial_{\vec{r}} \Psi =
      \partial_{\vec{r}}\cdot(\conj{\Psi}\Psi\vec{G}) -
      \conj{\Psi}\Psi(\partial_{\vec{r}}\cdot\vec{G})
  \end{aligned}
\end{equation}
In the Coulomb gauge, \(\nabla\cdot\vec{A}=0\), which means that the
second term above vanishes. Finally, we have used that
\(\conj{(\im\partial_{\vec{r}}\Psi)} =
-\im\partial_{\vec{r}}\conj{\Psi}\).

The \emph{probability current} is given by
\begin{equation}
  \label{eqn:current}
  \vec{j} =
  \frac{\im}{2}[
  (\tilde{\partial}_{\vec{r}}\conj{\Psi})\Psi -
  \conj{\Psi}\tilde{\partial}_{\vec{r}}\Psi
  ] -
  \frac{q_n}{m_n}\vec{A}(\vec{r},t)\conj{\Psi}\Psi,
\end{equation}
where we, as noted above, in the length gauge may choose
\(\vec{A}=0\); cf.\ equation~(115.4) in \cite{Landau1977quant}. Using
the divergence theorem, we can then write the probability in
\(\complementdomain\) as
\begin{equation}
  \label{eqn:probability}
  \matrixel{\Psi}{\complementmapping}{\Psi} =
  \int\diff{t}
  \int_\domain\diff{\vec{r}}
  \partial_{\vec{r}}\cdot\vec{j} =
  \int\diff{t}
  \oint_{\domainboundary}
  \vec{j}\cdot\diff{\vec{S}},
\end{equation}
i.e.\ the surface integral of the normal component of the current
flowing out through the boundary \(\domainboundary\), integrated over
time. As we shall see below, for differential spectra, we shall need
the \emph{transition currents}
\begin{equation}
  \label{eqn:transition-current}
  \vec{j}_{\Xi\Phi} \defd
  \frac{\im}{2}[
  (\tilde{\partial}_{\vec{r}}\conj{\Xi})\Phi -
  \conj{\Xi}\tilde{\partial}_{\vec{r}}\Phi
  ] +
  \im\conj{\Xi}\Phi\vec{G},
\end{equation}
where \(\Xi\) and \(\Phi\) are functions used in the expansion of the
wavefunction \(\Psi\), i.e.\ they all live in the same Hilbert
space. The transition currents can be derived in full analogy with
the ordinary ones, by making the substitution
\((\conj{\Psi},\Psi)\to(\conj{\Xi},\Phi)\) in
\eqref{eqn:complement-domain-probability}. We note that
\eqref{eqn:probability} \& \eqref{eqn:transition-current} are fully
equivalent with the commutator expression
\begin{equation}
  \label{eqn:commutator-formulation}
  \matrixel{\Xi}{\complementmapping}{\Phi} =
  \im
  \int\diff{t}
  \matrixel{\Xi}{\comm{\Hamiltonian}{\complementmapping}}{\Phi},
\end{equation}
which we are familiar with from the single-ionization case
\cite{Tao2012NJoP,Morales2016-isurf,Carlstroem2022tdcisI}. However, as
we shall see in the next section, we find the formulation of
\eqref{eqn:transition-current} preferable. Whenever we below encounter
a commutator such as the one appearing in
\eqref{eqn:commutator-formulation}, we must remember that it is to be
evaluated by means of a divergence of a current as in
\eqref{eqn:probability} \& \eqref{eqn:transition-current}. We note
that although \eqref{eqn:current} and \eqref{eqn:transition-current}
were derived assuming a non-relativistic Hamiltonian, a fully
analogous expression would result in the relativistic case, cf.\
appendix~\ref{app:relativistic-extension}.

\subsubsection{Example: The Line}
\label{sec:example-the-line}

\begin{figure}[htb]
  \centering
  \includegraphics[trim=30cm 30cm 30cm 30cm]{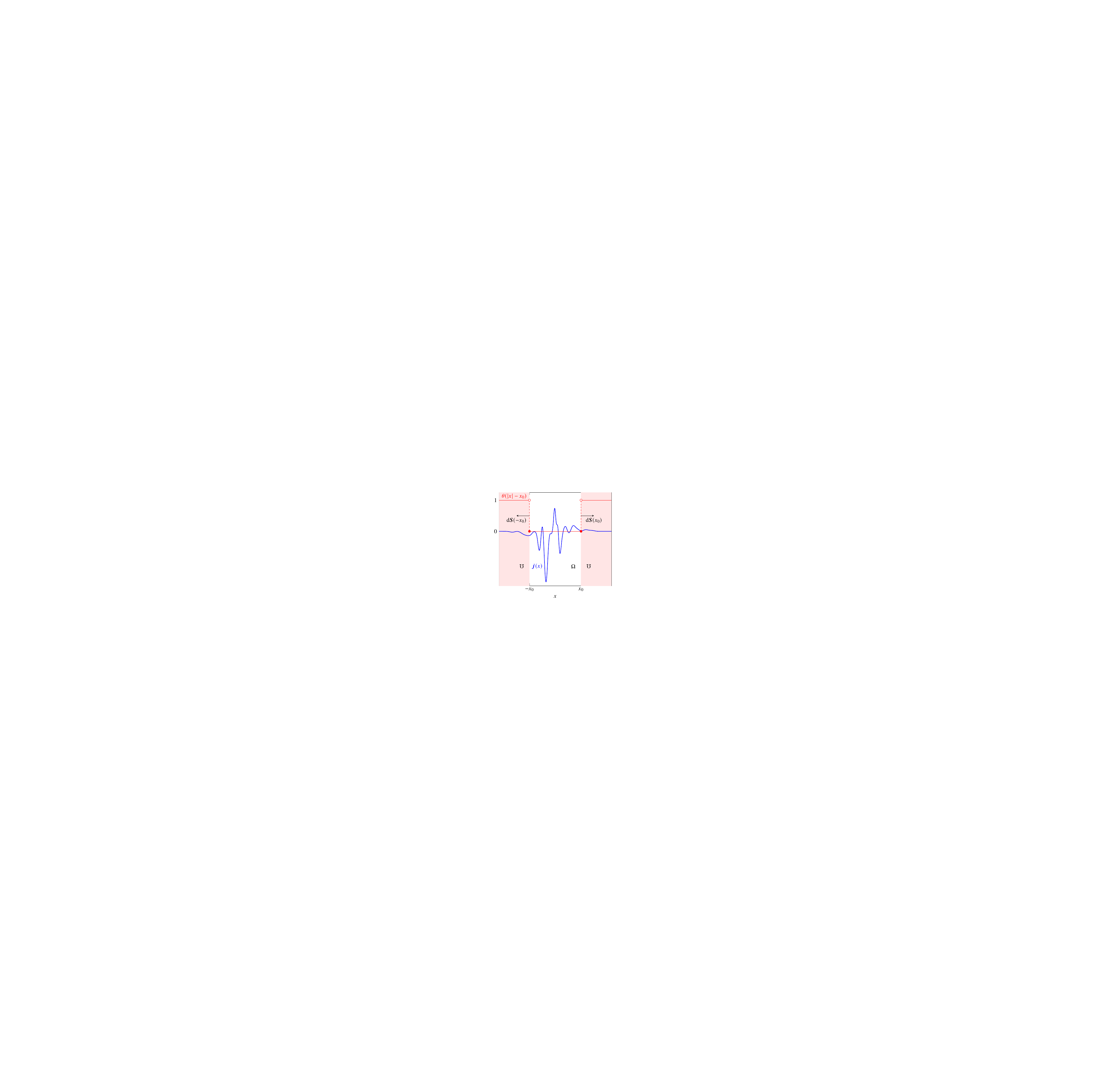}
  \caption{Partitioning of the line. We use open and closed circles to
    indicate that we have chosen the Heaviside function
    \eqref{eqn:heaviside} to be left-continuous. The blue line shows
    an example current density. The surface integration measure is a
    one-component unit vector:
    \(\diff{\vec{S}}(\pm x_0) = \bmat{\pm1}\), pointing away from the
    domain \(\domain\).}
  \label{fig:partitioning-of-the-line}
\end{figure}

Before continuing the derivation for general geometries, let us
consider the simple example of the line, which is exactly the
situation illustrated in figure~\ref{fig:partitioning-of-the-line}. In
this case, \(\domain\) is the line segment \([-x_0,x_0]\),
\(\complementdomain=\Real\setminus[-x_0,x_0]\), and the boundary
\(\domainboundary=\{-x_0,x_0\}\). The mapping \(\complementmapping\)
onto \(\complementdomain\) is given by
\eqref{eqn:one-dimensional-partitioning}. In this case, the
contribution to the probability in \(\complementdomain\) is trivially
given by
\begin{equation}
  \oint_{\domainboundary}
  \vec{j}\cdot\diff{\vec{S}} =
  \int_\Real\diff{x}
  j(x)[\delta(x_0)-\delta(-x_0)] =
  j(x_0) - j(-x_0).
\end{equation}

\subsection{Differential Geometry}
Actually computing integrals such as
\begin{equation}
  \oint_{\domainboundary}
  \vec{j}_{\Xi\Phi}\cdot\diff{\vec{S}}
\end{equation}
requires expressing \(\vec{j}_{\Xi\Phi}\) and the integration measure
\(\diff{\vec{S}}\equiv \vec{n}\diff{S}\) in the local coordinate
system. Since the submanifold \(\domainboundary\subset \Real^M\) is in
general non-Euclidean, we find the induced metric on the surface as
the pullback of the ambient metric according to
\begin{equation}
  \label{eqn:surface-metric}
  h_{ab}(\lambda) =
  \eta_{ij}[\vec{\Sigma}(\lambda)]
  \tensor{\Sigma}{^i_{,a}}
  \tensor{\Sigma}{^j_{,b}},
\end{equation}
where in common tensor notation,
\(\tensor{\Sigma}{^a_{,b}}\defd \partial \Sigma^a/\partial
\lambda^b\). Since our ambient space is \(\Real^M\), the ambient
metric is simply \(\eta_{ij}(\vec{r})=\delta_{ij}\), and the surface
metric simplifies to
\begin{equation}
  \tag{\ref{eqn:surface-metric}*}
  h_{ab}(\lambda) =
  \tensor{\Sigma}{_{i,a}}
  \tensor{\Sigma}{^i_{,b}}.
\end{equation}
The surface integration measure is then
\begin{equation}
  \diff{S} =
  \sqrt{\det h}
  \diff{\lambda^1}\wedge\diff{\lambda^2}\wedge \cdots \wedge \diff{\lambda^{M-1}}.
\end{equation}
Similarly, with the help of the Jacobian matrix
\(\mat{J}\in\Real^{M\times{}M}\) of the transformation
\((\lambda,\zeta)\to\vec{r}\),
\begin{equation}
  \label{eqn:jacobian}
  \mat{J} = \bmat{
    \frac{\partial r_1}{\partial\lambda}
    &
    \frac{\partial r_1}{\partial\zeta} \\
    \frac{\partial r_2}{\partial\lambda}
    &
    \frac{\partial r_2}{\partial\zeta} \\
    \vdots & \vdots \\
    \frac{\partial r_M}{\partial\lambda}
    &
    \frac{\partial r_M}{\partial\zeta}
  } \equiv
  \bmat{
    \frac{\partial\vec{r}}{\partial\lambda}
    &
    \frac{\partial\vec{r}}{\partial\zeta}
  } =
  \bmat{
    \frac{\partial\vec{\Sigma}(\lambda)}{\partial\lambda} +
    \zeta\frac{\partial\vec{n}(\lambda)}{\partial\lambda}
    &
    \vec{n}(\lambda)
  },
\end{equation}
we can find the local metric in a neighbourhood around
\(\lambda,\zeta\) as
\begin{equation}
  \label{eqn:metric}
  \begin{aligned}
    g_{ij}
    &\defd
      \partial_{\lambda_i}\vec{r}
      \cdot
      \partial_{\lambda_j}\vec{r} =
      \left[\frac{\partial\vec{\Sigma}}{\partial\lambda_i} +
      \zeta\frac{\partial\vec{n}}{\partial\lambda_i}\right]
      \cdot
      \left[\frac{\partial\vec{\Sigma}}{\partial\lambda_j} +
      \zeta\frac{\partial\vec{n}}{\partial\lambda_j}\right],\\
    g_{i\zeta}
    &\defd
      \partial_{\lambda_i}\vec{r}
      \cdot
      \partial_\zeta\vec{r}
      \left[\frac{\partial\vec{\Sigma}}{\partial\lambda_i} +
      \zeta\frac{\partial\vec{n}}{\partial\lambda_i}\right]
      \cdot
      \vec{n} = 0,
      \quad
      g_{\zeta j} = 0, \\
    g_{\zeta\zeta}
    &\defd
      \partial_\zeta\vec{r}
      \cdot
      \partial_\zeta\vec{r} =
      1,
  \end{aligned}
\end{equation}
where we have used that the normal \(\vec{n}\) is orthogonal to the
tangent plane spanned by \(\partial_\lambda\vec{r}\). This means our
metric is a block-diagonal matrix.

The gradient of a scalar field is transformed according to
\begin{equation}
  (\nabla f)^u =
  g^{uv}
  f_{,v},
\end{equation}
where \(g^{uv}\) is the matrix inverse of the metric \(g_{uv}\). We
thus find
\begin{equation}
  \label{eqn:surface-term-current}
  \oint_{\domainboundary}
  \vec{j}_{\Xi\Phi}\cdot\diff{\vec{S}} =
  \frac{\im}{2}
  \oint_{\domainboundary}
  g^{uv}[
  \tensor{\Xi}{^*_{,\tilde{u}}}\Phi -
  \conj{\Xi}\tensor{\Phi}{_{,\tilde{u}}} +
  2\conj{\Xi}\Phi G_u
  ]n_v\diff{S},
\end{equation}
where the subscript \(,\tilde{u}\) implies mass-weighting of the
differential along coordinate \(u\), in keeping with
\eqref{eqn:mass-weighted-differentials} and
\eqref{eqn:transition-current}. We note that all three terms in
\eqref{eqn:surface-term-current} are special cases of the surface
integral of an inner product with a vector field with the normal to
the surface:
\begin{equation}
  \label{eqn:surface-term-vector-field}
  \oint_{\domainboundary}
  g^{uv}
  V_un_v
  \diff{S}.
\end{equation}

It is difficult to make much more progress, without specifying the
domain \(\domain\). We will therefore now consider some common choices
and the specifics arising from the choice of computational domain.

\subsubsection{Spherical Symmetry}
Spherical symmetry has been mostly covered in earlier publications on
the topic \cite{Tao2012NJoP,Morales2016-isurf}. In \(\Real^3\)
spherical coordinates \(r,\theta,\phi\), with
\(x=r\sin\theta\cos\phi\), \(y=r\sin\theta\sin\phi\), and
\(z=r\cos\theta\), the local \(3\times3\) metric and its inverse are
given by
\begin{equation}
  \begin{aligned}
    (g)
    &=
      \operatorname{diag}
      \bmat{1 & r^2 & r^2\sin^2\theta} \\
    (g)^{-1}
    &=
      \operatorname{diag}
      \bmat{1 & r^{-2} & r^{-2}\sin^{-2}\theta}.
  \end{aligned}
\end{equation}
The induced metric on the surface \((h)\) is the lower-right
\(2\times2\) block of \((g)\):
\begin{equation}
  (h) =
  \operatorname{diag}
  \bmat{r^2 & r^2\sin^2\theta}.
\end{equation}
Thus \(\sqrt{\det h} = \sqrt{r^4\sin^2\theta} = r^2\sin\theta\), and
the surface integration measure is the familiar
\begin{equation}
  \diff{S} =
  r^2\sin\theta
  \diff{\theta}\wedge\diff{\phi}.
\end{equation}
Finally, a surface integral \eqref{eqn:surface-term-vector-field} of
the inner product of a vector field with the normal
\(\vec{n}=\uvec{r}\) is
\begin{equation}
  \label{eqn:surface-integral-spherical-symmetry}
  \oint_{\Sphere^2(r)}\!\!\!\!\!\!
  g^{uv}V_un_v\diff{S} =
  \oint_{\Sphere^2(r)}\!\!\!\!\!\!
  g^{u1}V_u\diff{S} =
  \oint_{\Sphere^2(r)}\!\!\!\!\!\!
  V^1\diff{S},
\end{equation}
where \(\Sphere^2(r)\) is the 2-sphere of radius \(r\). As we would
expect, we perform the surface integral over the radial component
(coordinate \(1\)) of the vector field
\cite{Tao2012NJoP,Morales2016-isurf}. We have kept the integration
measure \(\diff{S}\) symbolically in the last step, to not confuse the
square of the radius with the tensor index denoting coordinate.


\subsubsection{Cartesian Coordinates}
\label{sec:cartesian-coordinates}

In Cartesian coordinates \(\Real^N\), the natural choice for the
domain is the axis-aligned rectangular hypercuboid, whose faces span
\(\Real^{N-1}\subset\Real^N\). Both the ambient space and the manifold
are thus Euclidean, with a flat metric:
\begin{equation}
  \begin{aligned}
    (g) &= (g)^{-1} = \identity_N,
    &
    (h) &= \identity_{N-1},
  \end{aligned}
\end{equation}
and the surface integration measure is trivially
\begin{equation}
  \diff{S} = \diff{x^1}\wedge\diff{x^2}\wedge\cdots\wedge\diff{x^{N-1}}.
\end{equation}
The surface integral \eqref{eqn:surface-term-vector-field} simplifies
to a sum over \(2N\) hyperrectangles \(\Pi_{\pm i}\) of dimension
\(N-1\):
\begin{equation}
  \label{eqn:cartesian-surface-integrals}
  \begin{aligned}
    \oint_{\domainboundary}
    g^{uv}V_un_v\diff{S}
    &=
      \sum_{j=1}^N
      \oint_{\Pi_j}
      \!\!\!\diff{S}
      V^j -
      \oint_{\Pi_{-j}}
      \!\!\!\!\!\diff{S}
      V^j \\
    &\equiv
      \sum_{j=1}^N
      \left(
      \oint_{\Pi_j}
      \!\!\!\diff{S} -
      \oint_{\Pi_{-j}}
      \!\!\!\!\!\diff{S}
      \right)
      V^j,
  \end{aligned}
\end{equation}
where we have used that the normal of the plane \(\Pi_j\) is
\(\uvec{j}\), and that of the opposing plane \(\Pi_{-j}\) is
\(-\uvec{j}\).

Since the rectangular hypercuboid is not a smooth Riemannian manifold
everywhere, it is natural to ask what happens at the edges and
corners. However, the \(d\)-dimensional Hausdorff measure of any
object of dimension lower than \(d\) is zero; thus edges which are of
one dimension lower, and corners which are two dimensions lower, do
not contribute to the surface integral.

When the fragment wavefunctions are chosen as Volkov (super)waves (see
section~\ref{sec:volkov-waves} below), the vector fields in
\eqref{eqn:cartesian-surface-integrals} involve \(\exp(-\im k_nx^n)\)
and its derivative along the cardinal axes
\(\partial_{x^j}\exp(-\im k_ux^u) = -\im k^j\exp(-\im k_ux^u)\). We
immediately see that the surface integral
\begin{equation}
  \oint_{\Pi_j}
  W^j\exp(-\im k_ux^u)
\end{equation}
is the Fourier transform of \(U^j = W^j\exp(-\im k_jx^j)\) over all
\(N-1\) coordinates, except \(j\). If we employ a uniform
discretization of \(R^N\), the surface integrals are thus most
efficiently evaluated as \(2N\) \((N-1)\)-dimensional FFTs, for which
there are optimized implementations.

\subsubsection{The Line}

On the line \(\Real^1\), the situation is trivial: the natural
boundary is \(\Sphere^0(r)\), which only includes two points,
\(\{-r,r\}\), the metric is \((g)=\identity_1\), and the surface
integration measure is \(\diff{S}=\diff{x}\). The surface integral
\eqref{eqn:surface-term-vector-field} is simply a sum of two points:
\begin{equation}
  \label{eqn:line-surface-integrals}
  \oint_{\Sphere^0(r)}
  g^{uv}V_un_v\diff{S} =
  V(r) - V(-r),
\end{equation}
in agreement with our findings in section~\ref{sec:example-the-line}.

\subsection{Partitioning of Many-Body Space}
\label{sec:many-body-space-partitioning}
We now return to the common case of a domain constructed from
primitive domains \(\primitivedomain(i)\) for the various
fragments. Consider the case when wish to partition space such that
all bodies remain inside their respective domains
\(\primitivedomain(i)\), except a specified set \(\vec{v}\), which is
outside. The prototypical example would be \(N-v\) electrons remaining
bound within the ion, and \(v\) photoelectrons. We write the projector
onto this space as
\begin{equation}
  \label{eqn:many-body-projector}
  \mapping_{\vec{v}} \defd
  \bigoplus_{i\notin\vec{v}} \primitivemapping(i)
  \bigoplus_{j\in\vec{v}} [\identity-\primitivemapping(j)]
  \defd
  \primitivemapping_{\vec{v}}
  \oplus
  \primitivecomplementmapping_{\vec{v}},
\end{equation}
which projects all particles whose indices are \emph{not} in the set
\(\vec{v}\) onto the spaces \(\primitivedomain(i)\subset\Real^n\), and
those whose indices are onto the complement spaces
\(\Real^n\setminus\primitivedomain(i)\). We also introduce the
fragmentation projector 
\begin{equation}
  \label{eqn:fragmentation-projector}
  \fragmentationprojector
  \defd
  \sum_{\vec{v}}\mapping_{\vec{v}},
\end{equation}
which generates a decomposition of the wavefunction into various sets
\(\vec{v}\) of interest.

As a simple example of \eqref{eqn:many-body-projector}, the projector
applicable to single ionization would be
\begin{equation}
  \label{eqn:single-ionization-projector}
  \mapping_{\textrm{SI}} = \mapping_i,
\end{equation}
where, in a slight abuse of notation, we use the scalar subscript
\(i\) to mean the vector \(\vec{v}=\bmat{i}\) with the only element
being \(i\). The implied sum over \(i\) runs over all electrons, to
make the operator symmetric. Similarly, the double ionization
projector is given by
\begin{equation}
  \label{eqn:double-ionization-projector}
  \mapping_{\textrm{DI}} =
  \frac{1}{2}
  \mapping_{ij},
\end{equation}
under the implicit assumption that \(\mapping_{ii}=0\). The factor
\(\frac{1}{2}\) is necessary to avoid double-counting. Alternative
expressions, with the summations explicitly written out would be
\begin{equation}
  \tag{\ref{eqn:double-ionization-projector}*}
  \mapping_{\textrm{DI}} =
  \sum_{\substack{i\\j<i}}
  \mapping_{ij} =
  \frac{1}{2}
  \sum_{i\ne j}
  \mapping_{ij}.
\end{equation}
As a final example, the projector applicable for \emph{no} ionization
is then
\begin{equation}
  \label{eqn:no-ionization-projector}
  \mapping_{\textrm{NI}} =
  \mapping_{\emptyset} \equiv
  \mapping,
\end{equation}
in agreement with our earlier definition \eqref{eqn:mapping-operator}.

\subsection{Wavefunction \emph{Ansatz}}
Our \emph{Ansatz} is that the wavefunction may be written as a tensor
product
\begin{equation}
  \ket{\Psi(E)} =
  \bigotimes_i
  \ket{\psi_i(E_i)}
  \equiv
  \ket{\psi_1(E_1)\psi_2(E_2)\cdots\psi_n(E_n)},
\end{equation}
where the total energy is given as the sum of the energies of the
constituent parts:
\begin{equation}
  E = \sum_i E_i.
\end{equation}
We will typically need additional quantum numbers to unambiguously
identify the state \(\ket{\psi_i(E_i)}\).

Each and every constituent part obeys a subspace TDSE
\begin{equation}
  [\imdt - \hamiltonian_i(t)]\ket{\psi_i},
\end{equation}
as well as a corresponding eigenvalue equation
\begin{equation}
  \hamiltonian_i\ket{\psi_i} = E_i\ket{\psi_i},
\end{equation}
valid for time-independent Hamiltonians \(\hamiltonian_i\), i.e.\
typically when the external driving field has ended. The energy
\(E_i\) may fall into the discrete spectrum of the Hamiltonian
\(\hamiltonian_i\)
[\(\matrixel{\psi_i}{\identity-\primitivemapping(i)}{\psi_i}\simeq0\),
i.e.\ \(\ket{\psi_i}\) is localized within \(\primitivedomain(i)\)],
or it may fall into the continuous spectrum
[\(\matrixel{\psi_i}{\primitivemapping(i)}{\psi_i}\simeq0\), i.e.\
\(\ket{\psi_i}\) is asymptotically vanishing within
\(\primitivedomain(i)\)]. The total energy typically falls into the
continuous spectrum of the overall Hamiltonian
\(\Hamiltonian=\bigoplus_i\hamiltonian_i\), for the cases we are
commonly interested in. This means that at least one of the
constituent energies \(E_i\) must belong to the continuous
spectrum. Additionally, we note that there are many different possible
sets of the constituent energies \(\{E_i\}\) that add up to the same
total energy \(E\).

\subsection{Notation for Fragmentation Channels}
\label{sec:fragmentation-channels}

We first need to establish a notation for the fragmentation
channels. These correspond to what we measure, and is largely a choice
we make at our convenience. In the case there is a residual photoion
that we do not momentum-resolve, but only measure in terms of some
(discrete) quantum numbers (typically ionization potential and various
angular momenta), we by convention place this at the centre of the
coordinate system, and we label its various states by the collective
indices \(I,J,...\). Note that the photoion in states \(I\) and \(J\)
do not need to have the same charge or even be isomers of each
other. Fragmentation channels that share an ion state \(I\) are
collectively referred to as an \emph{ionization channel}. The
particles that we do momentum-resolve, we name \emph{fragments}. These
fragments may be elementary (electrons, protons, \&c.)\ or composite
particles (molecular fragments). Within a given fragmentation channel,
we may have different kinds of fragments simultaneously, and we by
convention require different kinds of fragments to be distinguishable,
and conversely particles belonging to the same kind of fragment are
thus indistinguishable. We use \(\varpi\) to label the fragment kind
(electron, proton, specific molecular fragment, ...). Each fragment
has a momentum \(\vec{k}\), a spin \(s\) and spin projection
\(\spin\).

A fragmentation channel thus consists of an optional photoion state
\(I\), and multiple fragment kinds. The \(i\)th fragment is labelled
\(\phi_i=(\varpi_i,s_i,w_i,k_i,\iota_i,S_i)\), where \(s_i\) is the
spin of one particle (\(1/2\) for electrons and protons, \(1\) for
deuterium, \&c.), \(w_i\) is occupancy of that fragment (number of
particles of this fragment kind),
\(k_i\defd\transpose{\bmat{\vec{k}_{i1}&\vec{k}_{i2}&\cdots&\vec{k}_{iw_i}}}\)
is the collective linear momentum of the fragment, where
\(\vec{k}_{ij}\) is the linear momentum of the \(j\)th particle of the
\(i\)th fragment, and \(\iota_i\) is an optional quantum number
labelling internal states of the fragment. Finally, \(S_i\) is either
\(\transpose{\bmat{\spin_1&\spin_2&\cdots&\spin_{w_i}}}\), if the
spins of the individual particles are resolved separately, or
\((S,M_S)_i\) with \(M_S=\sum_{j=1}^{w_i}\spin_j\) if the individual
spins \(s_i\) have been coupled to a good total spin \(S\) with
projection \(M_S\). The wavefunction for an individual fragment can
thus be written as
\begin{equation}
  \label{eqn:fragment-wavefunction}
  \ket{\phi_i} =
  \ket{(\varpi_is_i)^{w_i}k_i\iota_iS_i} =
  \left[
    \permsym_i
    \prod_{j=1}^{w_i}
    \ket{\varpi_is_i\vec{k}_{ij}\spin_j}
  \right]_{\iota_iS_i},
\end{equation}
where \(\permsym_i\) denotes the operator enforcing correct
permutation symmetry for particles of spin \(s_i\), i.e.
\begin{equation}
  \label{eqn:permutation-symmetry-operator}
  \permsym_i =
  \begin{cases}
    \antisym, & s_i\textrm{ is half-integer},\\
    \sym, & s_i\textrm{ is integer}.
  \end{cases}
\end{equation}
We note that the implicit summation/integration convention we employ
implies that the fragment wavefunction
\eqref{eqn:fragment-wavefunction} constitutes a wavepacket expanded
over a range of linear momenta \(k_i\), and an optional summation over
internal states \(\iota_i\).

The overall wavefunction for a fragmentation channel with \(v\)
fragment kinds thus becomes
\begin{equation}
  \label{eqn:fragmentation-channel-wavefunction}
  \ket{I\phi_1\phi_2\cdots\phi_vS}
  \equiv
  [\permsym\ket{I}\ket{\phi_1}\ket{\phi_2}\cdots\ket{\phi_v}]_S,
\end{equation}
where \(\permsym\) on the RHS enforces the correct permutation
symmetry between \(\ket{I}\) and \(\ket{\phi_i}\), since the fragment
kind \(\varpi_i\) may still exist in the photoion \(\ket{I}\) (the simplest
example is that of the outgoing electron in single ionization of
helium which needs to be antisymmetrized with respect to the remaining
electron in the cation \ch{He^+}). Since by construction, \(\varpi_i\) and
\(\varpi_j\) are distinguishable for \(i\ne j\) \footnote{This is strictly
  speaking not entirely correct; if \(\varpi_i\) denotes e.g.\ electrons
  and \(\varpi_j\) e.g.\ a molecular fragment, antisymmetrization
  \emph{should} actually be performed between \(\varpi_i\) and
  \(\varpi_j\). This is however expected to be a minor issue, since they
  would typically have rather different momenta, and their mutual
  exchange would contribute only negligibly.}, \(\permsym\) does not
permute particles between \(\ket*{\phi_i}\) and \(\ket*{\phi_j}\). Finally,
\(S\) is the overall spin of the fragmentation channel, which may
either be the sum of the spin projections of \(\ket{I}\) and the
various fragments \(\ket{\phi_i}\), or the total spin resulting from
coupling the individual spins, along with its projection. We also
introduce the compressed notation
\begin{equation}
  \tag{\ref{eqn:fragmentation-channel-wavefunction}*}
  \ket{I\vec{v}} \equiv
  \ket{I\phi_1\phi_2\cdots\phi_vS},
\end{equation}
consistent with the usage in
section~\ref{sec:many-body-space-partitioning}. The spin state \(S\)
of the wavefunction has been absorbed into the label for the state
\(I\) of the residual ion, since in practice we treat different spins
of the fragments correlated with a specific ion state as different
effective ionization channels. Different states \(\ket{I\vec{v}}\),
\(\ket{J\vec{v}}\) that share the same configuration \(\vec{v}\) of
fragment functions are said to belong to the same \emph{class of
  fragmentation channels}. We will see below that states within the
same class are coupled in a manner which is different compared to
states from different classes.

It is important to realize that the fragmentation channel wavefunction
\eqref{eqn:fragmentation-channel-wavefunction} lives in exactly the
same Hilbert space \(\HilbertSpace\) as does the wavefunction
\(\ket{\Psi}\). \eqref{eqn:fragmentation-channel-wavefunction} is
merely a convenient description of an asymptotic configuration of the
Hilbert space \(\HilbertSpace\), as mentioned in
section~\ref{sec:many-body-space-partitioning}. Specifically, the same
number of each kind of elementary particle must be present in both
\(\ket{\Psi}\) and \(\ket{I\vec{v}}\), i.e.\ no particles are created
or destroyed. If we take \(\varpi_i(\psi)\) to mean the number of
particles of kind \(\varpi_i\) which occur in the wavefunction
\(\ket{\psi}\), and
\(\varpi_i(I\vec{v}) \equiv \varpi_i(I) + \varpi_i(\phi_1) +
\varpi_i(\phi_2) + ... + \varpi_i(\phi_v)\), we thus require
\(\varpi_i(\Psi) = \varpi_i(I\vec{v}) = \varpi_i(I'\vec{v}')\).

Finally, we need to specify which TDSE is obeyed by the wavefunctions
\eqref{eqn:fragmentation-channel-wavefunction}. The crucial point of
the SURFF family of techniques is that the domain \(\domain\) is
chosen such that the Hamiltonian of the system in the exterior domain
\(\complementdomain\) is negligibly different from a scattering
Hamiltonian \(\scatteringHamiltonian(t)\), for which we know the
time-dependent eigenstates analytically. For ionization of atomic
hydrogen, the choice is the Volkov Hamiltonian
\begin{equation}
  \label{eqn:schroedinger-one-body-scattering-hamiltonian}
  \scatteringhamiltonian(t) =
  \frac{[\vecop{p}-q\vec{A}(\vec{r},t)]^2}{2m},
\end{equation}
typically in the dipole approximation
\(\vec{A}(\vec{r},t)\equiv\vec{A}(t)\), implying that we neglect the effect
of the nuclear Coulomb potential. See
section~\ref{sec:analytic-fragment-wavefunctions} below for a
discussion of the various approximations to \(\ket{\phi_i}\) that we may
employ in \eqref{eqn:fragment-wavefunction}. Here, we only sketch the
TDSE as
\begin{equation}
  \label{eqn:asymptotic-tdse}
  \begin{aligned}
    \imdt
    \ket{\Psi(t)}
    &\approx
      \scatteringHamiltonian(t)\ket{\Psi(t)}, \\
    \imdt
    \ket{I\vec{v}}
    &=
      \scatteringHamiltonian(t)\ket{I\vec{v}}, \\
    \iff
    -\imdt
    \bra{I\vec{v}}
    &=
      \bra{I\vec{v}}
      \scatteringHamiltonian(t),
  \end{aligned}
\end{equation}
where the scattering Hamiltonian is partitioned into an ionic part and
a fragment part:
\begin{equation}
  \label{eqn:scattering-hamiltonian}
  \scatteringHamiltonian(t) =
  \ionicHamiltonian(t) +
  \scatteringhamiltonian(t).
\end{equation}
The ionic states are chosen as the field-free ones, taken in the
interaction picture, leading to the trivial phase evolution
\begin{equation}
  \label{eqn:ionic-state-time-evolution}
  \ket{I(t)} =
  \ce^{-\im E_I(t-t_i)}
  \ket{I(t_i)},
\end{equation}
where \(E_I\) is the energy of the ionic state \(\ket{I}\), and
\(t_i\) is some reference time of our choosing. This time dependence
is assumed in the following, even if the time argument is suppressed
for brevity. The matrix elements of the ionic Hamiltonian between
these states are given by
\begin{equation}
  \label{eqn:ionic-hamiltonian-matrix-elements}
  \matrixel{I\vec{v}}{\ionicHamiltonian(t)}{I'\vec{v}'} =
  \delta_{\vec{v}\vec{v}'}
  \ce^{\im E_{II'}(t-t_i)}
  [E_I\delta_{II'} +
  \matrixel{I}{\laserinteraction(t)}{I'}],
\end{equation}
where it is important that the ionic states \(\ket{I}\) and
\(\ket{I'}\) belong to the same class \(\vec{v}\) of fragmentation
channels, \(E_{II'}\defd E_I-E_{I'}\) is the energy difference, and
\(\laserinteraction(t)\) is the interaction with the external field.

Similarly, the fragment state fulfils the fragment TDSE
\begin{equation}
  [\imdt-\scatteringhamiltonian(t)]\ket{\vec{v}} = 0,
\end{equation}
where the fragment Hamiltonian has the matrix elements
\begin{equation}
  \label{eqn:fragment-hamiltonian-matrix-elements}
  \matrixel*{\vec{v}'}{\scatteringhamiltonian(t)}{\vec{v}} =
  E_{\vec{v}}(t)\delta_{\vec{v}\vec{v}'}.
\end{equation}

The above formalism is valid in the general case. However, computing a
matrix element between two fragmentation channels
\begin{equation}
  \matrixel*{I\vec{v}}{\operator{A}}{I'\vec{v}'}
\end{equation}
or between a fragmentation channel and the wavefunction
\begin{equation}
  \matrixel*{I\vec{v}}{\operator{A}}{\Psi}
\end{equation}
for some \(N\)-body operator \(\operator{A}\) is decidedly a
non-trivial problem in the general case \cite{Bunker2006}. In
practice, the permutation symmetries will have to be worked out by
hand on a case-by-case basis. A class of matrix elements which
\emph{are} easy to compute, however, are those of the partitioning
operator:
\begin{equation}
  \begin{aligned}
    \matrixel*{I\vec{v}}{\mapping_{\vec{v}}}{\Psi}
    &=
      \bra*{\vec{v}}
      \primitivecomplementmapping_{\vec{v}}
      N
      \brakett*{I}{\Psi}
      \defd
      \matrixel*{\vec{v}}
      {\primitivecomplementmapping_{\vec{v}}}
      {\dyson{I}}
  \end{aligned}
\end{equation}
where we have used the decomposition
\eqref{eqn:many-body-projector}. The double right-angle bracket
\(\rAngle\) is used to indicate integration over both ionic and
fragment degrees of freedom; it is important to realize that
\(N\brakett{I}{\Psi}\) is \emph{not} a scalar. Indeed, in the last
step, we introduced the \emph{generalized Dyson orbital}
\(\ket*{\dyson{I}} \defd N\brakett{I}{\Psi}\), where \(N\) is the
appropriate normalization factor [e.g.\ if \(\ket{\Psi}\) is an
\(n\)-body function, and \(\ket{I}\) is an (\(n-1\))-body function,
\(N=\sqrt{n}\)]. The asymptotic TDSE obeyed by the generalized Dyson
orbital is
\begin{equation}
  \label{eqn:generalized-dyson-tdse}
  \begin{aligned}
    \imdt
    \ket*{\dyson{I}}
    &=
      \imdt
      N\brakett*{I}{\Psi}
      \approx
      -E_I\ket*{\dyson{I}} +
      N\matrixell*{I}{\scatteringHamiltonian(t)}{\Psi} \\
    &=
      -E_I\ket*{\dyson{I}} +
      N\matrixell*{I}
      {\ionicHamiltonian(t) + \scatteringhamiltonian(t)}
      {\Psi} \\
    &=
      [-E_I + \scatteringhamiltonian(t)]
      \ket*{\dyson{I}} +
      N\matrixell*{I}
      {\ionicHamiltonian(t)\identity}
      {\Psi} \\
    &=
      [-E_I + \scatteringhamiltonian(t)]
      \ket*{\dyson{I}} +
      \matrixel*{I}
      {\ionicHamiltonian(t)}
      {J}
      \ket*{\dyson{J}} \\
    &\defd
      \scatteringhamiltonian(t)
      \ket*{\dyson{I}} +
      \tilde{H}_{IJ}(t)
      \ket*{\dyson{J}}
  \end{aligned}
\end{equation}
where we have used equations~\eqref{eqn:asymptotic-tdse},
\eqref{eqn:scattering-hamiltonian},
\eqref{eqn:ionic-state-time-evolution}, and
\eqref{eqn:ionic-hamiltonian-matrix-elements}, and inserted a
resolution-of-identity in terms of ionic states
\begin{equation}
  \identity = \ketbra{J}{J},
\end{equation}
and introduced the ion channel coupling matrix \(\tilde{\mat{H}}(t)\)
with matrix elements \cite[cf.\ equation~(9)
of][]{Carlstroem2022tdcisI}
\begin{equation}
  \label{eqn:ion-channel-coupling-matrix}
  \tilde{H}_{IJ}(t)
  \defd
  \ce^{\im E_{IJ}(t-t_i)}
  \matrixel{I}{\laserinteraction(t)}{J}.
\end{equation}

\begin{figure}[htb]
  \centering
  \includegraphics[trim=30cm 30cm 30cm 30cm]{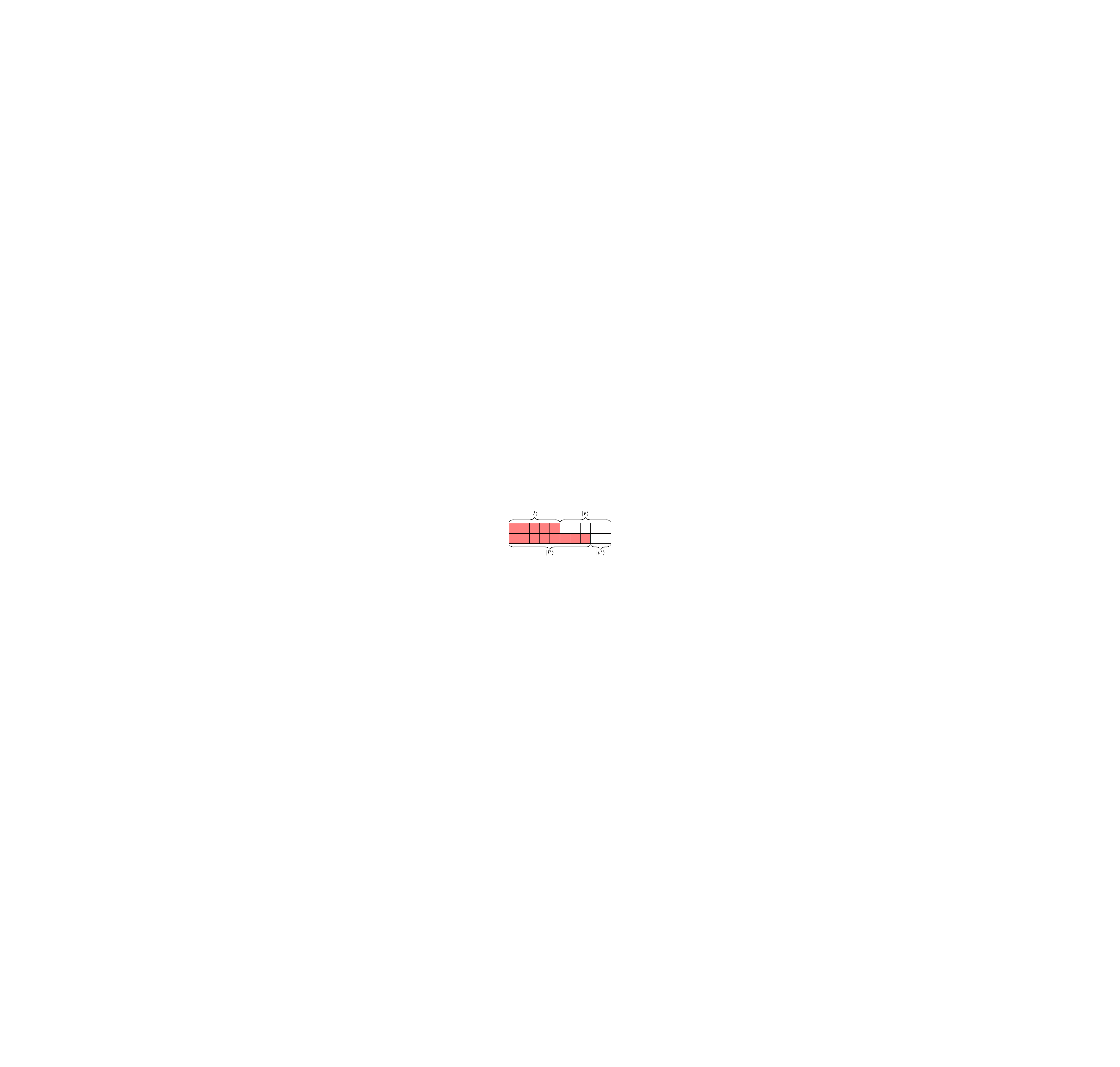}
  \caption{Pictorial illustration of the fragmentation channel
    overlap
    \(\matrixel*{I\vec{v}}{\mapping_{\vec{v}}}{I'\vec{v}'}\). The
    shaded cells indicate coordinates occupied by the ionic states
    (inside \(\domain\)), and unshaded cells indicate coordinates
    occupied by fragment states (outside \(\domain\)).}
  \label{fig:fragmentation-channel-overlap}
\end{figure}
Similarly, the matrix element of the partitioning operator between two
different fragmentation channels (illustrated in
figure~\ref{fig:fragmentation-channel-overlap})
\begin{equation}
  \label{eqn:dyson-tensor}
  \begin{aligned}
    \matrixel*{I\vec{v}}{\mapping_{\vec{v}}}{I'\vec{v}'}
    &=
      \bra*{\vec{v}}
      {\primitivecomplementmapping_{\vec{v}}}
      N
      \brakett*{I}
      {I'\vec{v}'}
      \defd
      \matrixel*{\vec{v}}
      {\dysontensor{I}{I'}}
      {\vec{v}'}
  \end{aligned}
\end{equation}
where \(\dysontensor{I}{I'}\) is not even a generalized Dyson orbital
anymore, but an object we name \emph{Dyson tensor}, since it behaves
like an operator. Finally, we note that
\begin{equation}
  \fragmentationprojector\ket{I\vec{v}}
  \equiv
  \mapping_{\vec{v}}\ket{I\vec{v}},
\end{equation}
such that 
\begin{equation}
  \begin{aligned}
    \matrixel{I\vec{v}}{\fragmentationprojector}{\Psi}
    &\equiv
      \matrixel{I\vec{v}}{\mapping_{\vec{v}}}{\Psi}, \\
      \matrixel{I\vec{v}}{\fragmentationprojector}{I'\vec{v}'}
    &\equiv
      \matrixel{I\vec{v}}{\mapping_{\vec{v}}}{I'\vec{v}'} \equiv
      \matrixel{I\vec{v}}{\mapping_{\vec{v}'}}{I'\vec{v}'}.
  \end{aligned}
\end{equation}

\subsubsection{Example: Single and Double Ionization of \ch{He}}
\label{sec:fragmentation-channels-He}
For atoms, it is natural to assume that the cation remains
approximately stationary, and we only momentum-resolve the
photoelectrons. For single ionization of helium, the fragmentation
channels are thus
\begin{equation}
  \begin{aligned}
    \ket*{\ch{He^{2+}}\;n\ell(^2L)\;\vec{k}S}
    &=
      [\antisym
      \ket*{\ch{He^{2+}}}
      \ket*{n\ell\;{^2L}}
      \ket{\vec{k}}]_S \\
    &=
      \ket*{\ch{He^{2+}}}
      [\antisym
      \ket*{n\ell\;{^2L}}
      \ket{\vec{k}}]_S,
  \end{aligned}
\end{equation}
where \(L=\ell\) by necessity and \(S\) is the total spin formed from
coupling the spin \(1/2\) of the bound electron with the spin \(1/2\)
\(\implies S=0,1\), i.e.\ only singlets and triplets are
possible. Equivalently, we could have chosen either a \(jj\)-coupling
or intermediate coupling (\(LSJ\)) description of the core, and
coupled the spin of the photoelectron to the \(J\) of the photoion,
but the philosophy is the same. The antisymmetrization operator
\(\antisym\) only acts on the electronic degrees of freedom.

For double ionization of helium, there is only one fragmentation
channel:
\begin{equation}
  \ket*{\ch{He^{2+}}\;\vec{k}_1\vec{k}_2S} =
  \ket*{\ch{He^{2+}}\;}
  [\antisym
  \ket{\vec{k}_1}
  \ket{\vec{k}_2}]_S,
\end{equation}
and the dication \ch{He^{2+}} is by necessity in its only state, the
ground state (unless we wish to consider excited states of the
nucleus).

\subsubsection{Example: Single and Double Ionization or Disassociation
  of \ch{H_2}}
For single ionization of \ch{H_2}, the ionization-only fragmentation
channels we need to consider is \(\electron+\ch{H_2^+}\), the
wavefunction of which can be written as
\begin{equation}
  \ket{I\vec{k}S} =
  [\antisym\ket{I}
  \ket{\vec{k}}]_S
\end{equation}
where \(\ket{I}\) is a (possibly excited) state of the cation
\ch{H_2^+}.

We also have \(\electron+\ch{H}+\ch{H^+}\), i.e.\ single ionization
followed by disassociation. In this case all three fragments are
distinguishable, and the wavefunction is simply a product:
\begin{equation}
  \ket*{
    \vec{k}_{\electron}\;
    \ch{H}\gamma_{\ch{H}}\vec{k}_{\ch{H}}\;
    \ch{H^+}\vec{k}_{\ch{H^+}}\;
    S
  } =
  [
  \ket*{\vec{k}_{\electron}}
  \ket*{\ch{H}\gamma_{\ch{H}}\vec{k}_{\ch{H}}}
  \ket*{\ch{H^+}\vec{k}_{\ch{H^+}}}
  ]_S,
\end{equation}
where \(\gamma_{\ch{H}}\) is an extra quantum number to disambiguate
excited states of the \ch{H} fragment.

Double ionization is in some sense the simplest case, since the only
fragmentation channel is \(2\electron+2\ch{H^+}\).

More intricate is the photoexcitation followed by disassociation,
i.e.\ the fragmentation channel \(2\ch{H}\), where the particles are
may or may not be distinguishable, depending on the internal structure
of the individual particles (e.g.\ one may be in the electronic ground
state whereas the other one may be excited). We thus have to
explicitly list the fragmentation channels of interest, and apply
permutation symmetrization when appropriate.


\subsection{Equations of Motion for Fragmentation Amplitudes}
We have now reached the point where we can apply the formalism
developed in the previous sections to the problem at hand, namely the
derivation of a hierarchy of equations describing the evolution of the
system in terms of the fragmentation channels. This is highly
reminiscent of the single ionization, multiple ionization-channels
case derived in \cite{Carlstroem2022tdcisI}, here generalized to
multiple fragmentation. The basic idea is that the division of space
as enforced by the projection operators
\eqref{eqn:fragmentation-projector} defined in
section~\ref{sec:space-partitioning} allows us to decompose the
wavefunction into the fragmentation channel wavefunctions described in
section~\ref{sec:fragmentation-channels}:
\begin{equation}
  \label{eqn:asymptotic-ansatz}
  \fragmentationprojector
  \ket{\Psi(t)} =
  a_{I\vec{v}}(t)
  \ket{I\vec{v}},
\end{equation}
where it is understood that the \(v\) indices in the set \(\vec{v}\)
pertain to the fragments
\(\ket{\phi_1}\ket{\phi_2}\cdots\ket{\phi_v}\), and the indices not in this set to
the residual ion \(\ket{I}\). The hierarchy arises from specifying a
number of different sets \(\{\vec{v}_i\}\), each set \(\vec{v}_i\)
corresponding a class of fragmentation channels (single ionization,
double ionization, disassociation, \&c.), and by time-differentiating
the fragmentation channel coefficient \(a_{I\vec{v}}(t)\), we
arrive at a coupled set of equations describing the time evolution in
terms of the fragmentation channel wavefunctions, in a fashion very
similar to single-ionization case described in
\cite{Carlstroem2022tdcisI}. There are however extra terms arising
from the coupling between different classes in the hierarchy, as we
will see below.

As is commonly observed, it is enough to consider only one term in
e.g.\ \eqref{eqn:single-ionization-projector} when applied to the
wavefunction \(\ket{\Psi(t)}\); the operator being symmetric and the
wavefunction properly permutation-symmetrized will ensure that we may
choose any term, say the first, and get the correct result. For the
double-ionization projector \eqref{eqn:double-ionization-projector},
it is enough to choose any two terms, say the first two, to arrive at
the correct result.

\subsubsection{Equations of Motion}

We will now derive the general equations of motion (EOMs) for the
fragmentation amplitudes. In
section~\ref{sec:fragmentation-channels-He-EOMs} below, we illustrate
the EOMs by specializing to the concrete example of single and double
ionization of \ch{He}.

We begin by rewriting the asymptotic \emph{Ansatz}
\eqref{eqn:asymptotic-ansatz} as an inner product of two vectors
\begin{equation}
  \label{eqn:asymptotic-ansatz-vector-form}
  \tag{\ref{eqn:asymptotic-ansatz}*}
  \fragmentationprojector
  \ket{\Psi(t)} =
  \transpose{\vec{\phi}}
  \vec{a}(t)
\end{equation}
where the row vector of fragmentation channels is given by
\begin{equation}
  \transpose{\vec{\phi}}
  \defd
  \bmat{
    \ket{I\vec{v}} & \ket{J\vec{v}} & \cdots &
    \ket{I'\vec{v}'} & \ket{J'\vec{v}'} & \cdots
  },
\end{equation}
their time-dependent expansion coefficients by
\begin{equation}
  \label{eqn:fragmention-amplitudes}
  \transpose{\vec{a}}(t)
  \defd
  \bmat{
    a_{I\vec{v}}(t) & a_{J\vec{v}}(t) & \cdots &
    a_{I'\vec{v}'}(t) & a_{J'\vec{v}'}(t) & \cdots
  },
\end{equation}
and we have ordered the elements by the fragmentation classes
\(\vec{v}\), \(\vec{v}'\), .... We may turn this into an equation
system by multiplying \eqref{eqn:asymptotic-ansatz-vector-form} from
the left by \(\conj{\vec{\phi}}\fragmentationprojector\), which is a
column vector of duals of the fragmentation channels (suppressing the
time dependence for brevity):
\begin{equation}
  \label{eqn:fragmentation-channels-equation-system}
  \vec{\Psi} = \mat{B}\vec{a}.
\end{equation}
The vector \(\vec{\Psi}\) on the left-hand side consists of overlaps
between the various fragmentation channels and the wavefunction
\(\ket{\Psi}\):
\begin{equation}
  \label{eqn:wavefunction-fragmentation-channel-overlap}
  \vec{\Psi} \defd
  \conj{\vec{\phi}}
  \fragmentationprojector
  \ket{\Psi}
  \equiv
  \bmat{
    \matrixel{I\vec{v}}{\fragmentationprojector}{\Psi} \\
    \matrixel{J\vec{v}}{\fragmentationprojector}{\Psi} \\
    \vdots \\
    \matrixel{I'\vec{v}'}{\fragmentationprojector}{\Psi} \\
    \matrixel{J'\vec{v}'}{\fragmentationprojector}{\Psi} \\
    \vdots
  }.
\end{equation}
Similarly, the matrix \(\mat{B}\) on the right-hand side contains
overlaps between fragmentation channels:
\begin{equation}
  \label{eqn:fragmentation-channel-overlap}
  \mat{B} \defd
  \conj{\vec{\phi}}
  \fragmentationprojector
  \transpose{\vec{\phi}}
  \equiv
  \bmat{
    1 & 0 & \cdots & B_{I\vec{v}}^{I'\vec{v}'} &
    B_{I\vec{v}}^{J'\vec{v}'} & \cdots \\[1ex]
    0 & 1 & \cdots & B_{J\vec{v}}^{I'\vec{v}'} &
    B_{J\vec{v}}^{J'\vec{v}'} & \cdots \\[1ex]
    \vdots & \vdots & \ddots & \vdots & \vdots & \ddots \\[1ex]
    B_{I'\vec{v}'}^{I\vec{v}} &
    B_{I'\vec{v}'}^{J\vec{v}} & \cdots & 1 & 0 &
    \cdots \\[1ex]
    B_{J'\vec{v}'}^{I\vec{v}} &
    B_{J'\vec{v}'}^{J\vec{v}} & \cdots & 0 & 1 &
    \cdots \\[1ex]
    \vdots & \vdots & \ddots & \vdots & \vdots & \ddots
  },
\end{equation}
where \(B_a^b\defd\matrixel{a}{\fragmentationprojector}{b}\). In
writing \eqref{eqn:fragmentation-channel-overlap}, we have made use of
the fact that the overlaps within a given fragmentation class
\(\vec{v}\) are orthonormal:
\begin{equation}
  \matrixel{I\vec{v}}{\fragmentationprojector}{I'\vec{v}} = \delta_{II'},
\end{equation}
by virtue of \(\ket{I}\), \(\ket{I'}\) being eigenstates of the same
ionic Hamiltonian \eqref{eqn:ionic-hamiltonian-matrix-elements}, and
the various fragments \(\ket{\phi_i}\) are always orthogonal to the
ion states \(\ket{I}\) by construction (the so-called strong
orthogonality assumption
\cite{Martin1976,Pickup1977,Oehrn1981,Patchkovskii2006}). The
off-diagonal blocks contain contributions from different fragmentation
classes \(\vec{v}\), \(\vec{v}'\), whose overlaps do not vanish
\emph{a priori}. We will see below that they play an important rôle in
resolving the time-ordering of the fragmentation process.

The EOMs are found by time-differentiating
\eqref{eqn:fragmentation-channels-equation-system}:
\begin{equation}
  \label{eqn:fragmentation-channel-eoms}
  \partial_t\vec{\Psi} =
  \dot{\mat{B}}\vec{a} +
  \mat{B}\partial_t\vec{a}.
\end{equation}
The time derivative of
\eqref{eqn:wavefunction-fragmentation-channel-overlap}, in full
analogy with the single-ionization case \cite{Carlstroem2022tdcisI},
yields the \emph{source term} \(\vec{S}(t)\) and the ion channel
couplings:
\begin{equation}
  \label{eqn:source-term-vector-form}
  -\imdt\vec{\Psi}(t) =
  \vec{S}(t) -
  \tilde{\mat{H}}(t)\vec{a}(t).
\end{equation}
For clarity, one component of \eqref{eqn:source-term-vector-form} is
found as
\begin{equation}
  \label{eqn:source-term}
  \begin{aligned}
    -\imdt
      \matrixel*{I\vec{v}}{\mapping}{\Psi(t)}
    &\equiv
    -\imdt
      \matrixel*{\vec{v}}
      {\primitivecomplementmapping_{\vec{v}}}
      {\dyson{I}} \\
    &=
      \underbrace{\matrixel*{\vec{v}}
      {\comm{\scatteringhamiltonian(t)}
      {\primitivecomplementmapping_{\vec{v}}}}
      {\dyson{I}}}_{\defd S_{I\vec{v}}(t)} -
      \tilde{H}_{IJ}(t)
      a_{J\vec{v}}(t),
  \end{aligned}
\end{equation}
where we have used \eqref{eqn:generalized-dyson-tdse}, and the ion
channel coupling matrix \(\tilde{\mat{H}}(t)\) is given by
\eqref{eqn:ion-channel-coupling-matrix}; it is important to note that
it only couples between ion channels within the same fragmentation
class \(\vec{v}\).

The right-hand side of \eqref{eqn:fragmentation-channel-eoms} contains
the fragmentation channel overlap matrix \(\mat{B}\)
\eqref{eqn:fragmentation-channel-overlap} and its time derivative.
Even in light of \eqref{eqn:dyson-tensor}, evaluating the volume
integrals \(\matrixel*{I\vec{v}}{\fragmentationprojector}{I'\vec{v}'}\) appearing in
the off-diagonal blocks of \(\mat{B}\) may prove to be a daunting
task. We therefore employ the trick of differentiating, and then
integrating in time:
\begin{equation}
  \mat{B}(t) =
  \int_{t_i}^t\diff{t}
  \dot{\mat{B}},
\end{equation}
where we require the initial condition
\begin{equation}
  \label{eqn:fragmentation-channel-overlap-initial-condition}
  \begin{aligned}
    \mat{B}(t)
    &=
      \identity,
    &
      t &\le t_i.
  \end{aligned}
\end{equation}
We are then left with the task of evaluating the
\emph{transfer term}
\begin{equation}
  \label{eqn:transfer-term}
  T_{I\vec{v}}^{I'\vec{v}'}(t)
  \defd
  -\imdt
  \matrixel*{I\vec{v}}{\fragmentationprojector}{I'\vec{v}'} =
  \matrixel*{I\vec{v}}{
    \comm*{\scatteringHamiltonian(t)}{\fragmentationprojector}
  }{I'\vec{v}'},
\end{equation}
named thus, since it transfers amplitude from the fragmentation
channel \(\ket*{I'\vec{v}'}\) to \(\ket*{I\vec{v}}\). Again, we have
traded the volume integral for an integral over surface and time,
exactly the same way we do in ordinary tSURFF:
\begin{equation}
  \dot{B}_{I\vec{v}}^{I'\vec{v}'} =
  \im
  (1-\delta_{\vec{v}\vec{v}'})
  T_{I\vec{v}}^{I'\vec{v}'}(t),
\end{equation}
and thus
\begin{equation}
  \tag{\ref{eqn:fragmentation-channel-overlap}*}
  \label{eqn:fragmentation-channel-overlap-matrix}
  B_{I\vec{v}}^{I'\vec{v}'}
  \equiv
  \delta_{II'}
  \delta_{\vec{v}\vec{v}'} +
  \im
  (1-\delta_{\vec{v}\vec{v}'})
  \left[
    \int_{t_i}^{t}\diff{t}
    T_{I\vec{v}}^{I'\vec{v}'}(t)
  \right].
\end{equation}

In both \eqref{eqn:source-term} and \eqref{eqn:transfer-term}, we have
used \eqref{eqn:asymptotic-tdse} to arrive at the commutator
expression we are familiar with from the single-ionization case
\cite{Tao2012NJoP,Morales2016-isurf,Carlstroem2022tdcisI}, which we
remember from section~\ref{sec:probability-in-complement-domain} is to
be evaluated as the divergence of a current
\eqref{eqn:transition-current}:
\begin{equation}
  \label{eqn:source-term-divergence-form}
  \begin{aligned}
    S_{I\vec{v}}(t) =
    -\im
    \oint_{\domainboundary}
    \vec{j}_{\vec{v}\tilde{I}}\cdot\diff{\vec{S}}
    &=
      \frac{1}{2}
      \oint_{\domainboundary}
      g^{uv}[\conj{V}_{,\tilde{u}}\tilde{I} -
      \conj{V}\tilde{I}_{,\tilde{u}}]n_v\diff{S} \\
    &\hphantom{=}+
      \oint_{\domainboundary}
      \conj{V}\tilde{I}g^{uv}G_un_v\diff{S}
  \end{aligned}
\end{equation}
where \(V\) is the scalar field associated with the scattering state
\(\ket{\vec{v}}\), and similarly, \(\tilde{I}\) is the scalar field
associated with the Dyson orbital \(\ket*{\tilde{I}}\). Similarly,
\begin{equation}
  \begin{aligned}
    T_{I\vec{v}}^{I'\vec{v}'}(t)
    &=
      -\im
      \oint_{\domainboundary}
      \vec{j}_{I\vec{v},I'\vec{v}'} \\
    &=
      \frac{1}{2}
      \oint_{\domainboundary}
      g^{uv}[\conj{(\tilde{I}V)}_{,\tilde{u}}(\tilde{I}'V') -
      \conj{(\tilde{I}V)}(\tilde{I}'V')_{,\tilde{u}}]n_v\diff{S} \\
    &\hphantom{=}+
      \oint_{\domainboundary}
      \conj{(\tilde{I}V)}(\tilde{I}'V')g^{uv}G_un_v\diff{S}.
  \end{aligned}
\end{equation}

Putting all this together, we arrive at the final expression for the
amplitude EOMs
\begin{equation}
  \label{eqn:fragmentation-channel-eoms-matrix-form}
  \tag{\ref{eqn:fragmentation-channel-eoms}*}
  [\mat{B}
  \partial_t +
  \im\mat{H}(t)]
  \vec{a}(t) =
  \im\vec{S}(t)
\end{equation}
which is a \emph{generalized}, \emph{inhomogeneous} TDSE for the
amplitude vector \(\vec{a}(t)\) \eqref{eqn:fragmention-amplitudes}
which plays the part of the wavefunction. The Hamiltonian
\(\mat{H}(t) = \tilde{\mat{H}}(t)-\im\dot{\mat{B}}\) of this TDSE has
matrix elements given by
\begin{equation}
  H_{I\vec{v}}^{I'\vec{v}'}(t)
  \defd
  \delta_{\vec{v}\vec{v}'}
  \tilde{H}_{II'}(t) +
  (1-\delta_{\vec{v}\vec{v}'})
  T_{I\vec{v}}^{I'\vec{v}'}(t),
\end{equation}
which contains the laser-driven transitions between ionic states
within the same fragmentation class \(\vec{v}\) in the first term
[cf.\ equation~\eqref{eqn:ion-channel-coupling-matrix}], and transfers
between different fragmentation classes in the second term.

The TDSE is \emph{generalized} due to the non-diagonal overlap matrix
\(\mat{B}\) \eqref{eqn:fragmentation-channel-overlap-matrix}. In the
case of only one class of fragmentation channels, e.g.\ single
ionization, or (as discussed in
section~\ref{sec:transfer-term-interpretation} below) if the ionic
states are completely localized within \(\domain\), the overlap matrix
\(\mat{B}\) naturally reduces to the identity matrix, and we recover
equation~(9\('\)) of
\cite{Carlstroem2022tdcisI}. 
Furthermore, the TDSE is \emph{inhomogeneous} due to the source term
on the right-hand side, containing the vector
\(\vec{S}(t) \defd \bmat{S_{I\vec{v}}(t)}\).

\subsubsection{Time-Stepping Scheme}
We may devise a time-stepping scheme which is globally accurate to
second order in the time step \(\timestep\), by discretizing
\eqref{eqn:fragmentation-channel-eoms-matrix-form} in time and
applying the trapezoidal rule:
\begin{equation}
  \frac{\mat{B}_{j+1}\vec{a}_{j+1} - \mat{B}_j\vec{a}_j}{\timestep} +
  \frac{\im}{2}
  (
  \mat{H}_{j+1}
  \vec{a}_{j+1} +
  \mat{H}_j
  \vec{a}_j
  ) =
  \frac{\im}{2}
  (\vec{S}_{j+1} + \vec{S}_j),
\end{equation}
which we may rearrange into a form reminiscent of the Crank--Nicolson
scheme:
\begin{equation}
  \label{eqn:time-stepping-scheme}
  \left(
    \mat{B}_{j+1} +
    \frac{\im\timestep}{2}
    \mat{H}_{j+1}
  \right)
  \vec{a}_{j+1} =
  \left(
    \mat{B}_j -
    \frac{\im\timestep}{2}
    \mat{H}_j
  \right)
  \vec{a}_j +
  \frac{\im\timestep}{2}
  (\vec{S}_{j+1} + \vec{S}_j).
\end{equation}
Our time grid is \(t_j = t_i + (j-1)\timestep\), where \(t_i\) is the
initial time, \(\timestep\) the length of the time step, and \(j\) the
index of the time step. \(\vec{S}_1 = 0\) by virtue of the initial
wavefunction being confined,
\(\braket*{\vec{r}}{\Psi(t_i)} \in \Omega\). Similarly, by the initial
condition \eqref{eqn:fragmentation-channel-overlap-initial-condition},
we have \(\mat{B}_1\equiv\identity\). We accomplish one step from
\(j\) to \(j+1\) using \eqref{eqn:time-stepping-scheme} by evaluating
the source term \(\vec{S}_{j+1}\) [which requires knowledge of the
wavefunction \(\ket*{\Psi(t_{j+1})}\)], and the surface integral in
the transfer term \(T_{I\vec{v}}^{I'\vec{v}'}(t_{j+1})\). The overlap
matrix is updated using the trapezoidal rule as
\(\mat{B}_{j+1} = \mat{B}_j + \im\timestep\tilde{\mat{T}}\), where
\begin{equation}
  \tilde{T}_{I\vec{v}}^{I'\vec{v}'} =
  (1-\delta_{\vec{v}\vec{v}'})
  \frac{T_{I\vec{v}}^{I'\vec{v}'}(t_{j+1}) +
    T_{I\vec{v}}^{I'\vec{v}'}(t_j)}{2}.
\end{equation}

\subsubsection{Terminating the Hierarchy; Closure Relations}
The hierarchical EOMs can very quickly become overwhelming, especially
if allow arbitrary levels of fragmentation, since all classes of
fragmentation channels communicate through the presence of
\(\matrixel*{I\vec{v}}{\fragmentationprojector}{I'\vec{v}'}\) in
\eqref{eqn:transfer-term}. One way of making this tractable is to
somewhat arbitrarily exclude all couplings but a select few that are
deemed of physical importance, this closes the hierarchy, and we label
the selection criteria \emph{closure relations}. The overlap is
pictorially illustrated in
figure~\ref{fig:fragmentation-channel-overlap}. If \(v\) is the number
of coordinates occupied by the fragment state \(\ket*{\vec{v}}\) and
similarly \(v'\) the number of coordinates occupied by the fragment
state \(\ket*{\vec{v}'}\), then the difference \(\Delta v=v-v'\) is
the number of extra fragments that have to be expelled for the
\(\matrixel*{I\vec{v}}{\fragmentationprojector}{I'\vec{v}'}\) to not
vanish. One closure relation could be the requirement \(\Delta v>0\),
which corresponds to neglecting the possibility of
recombination. Another, stricter, closure relation would be
\(q\ge\Delta v>0\), additionally neglecting non-sequential
fragmentation involving more than \(q\) fragments.

\subsubsection{Interpretation of the Transfer Terms}
\label{sec:transfer-term-interpretation}

From figure~\ref{fig:fragmentation-channel-overlap}, we infer that the
fragmentation channel overlap
\(\matrixel*{I\vec{v}}{\fragmentationprojector}{I'\vec{v}'}\) [and
hence also the transfer term \eqref{eqn:transfer-term}] vanishes, if
the ionic states \(\ket{I},\ket{I'}\) are completely localized within
\(\domain\). Sequential fragmentation, which is accounted for using
the transfer term \eqref{eqn:transfer-term} would thus appear to not
be covered by this approach; this is however not the case. If in the
first stage, fragmentation occurs into the fragmentation channel
\(\ket{I\vec{v}}\), with the ionic state \(\ket{I}\) localized
entirely within \(\domain\), sequential fragmentation into
\(\ket{I'\vec{v}'}\) occurs through by first coupling
\(\ket{I\vec{v}}\) to a set of higher excited states
\(\{\ket{I^*\vec{v}}\}\) within the same fragmentation class
\(\vec{v}\) through the ion coupling matrix
\eqref{eqn:ion-channel-coupling-matrix}. When a sufficiently highly
excited ionic state \(\ket{I^*}\) has been reached, which is non-zero
at the boundary \(\domainboundary\) (see e.g.\ right panel of
figure~\ref{fig:he-energy-level-scheme}, below), the
fragmentation channel overlap
\(\matrixel*{I^*\vec{v}}{\fragmentationprojector}{I'\vec{v}'}\) is
non-zero, and sequential fragmentation may occur. If the expansion in
terms of ionic states is truncated, such that only states localized
within \(\domain\) are included, the off-diagonal blocks of
\eqref{eqn:fragmentation-channel-overlap} thus vanish. As a
consequence, the EOMs simplify, since the overlap matrix
\eqref{eqn:fragmentation-channel-overlap-matrix} reduces to the
identity matrix, and the generalized, inhomogeneous TDSE
\eqref{eqn:fragmentation-channel-eoms-matrix-form} reduces to the
ordinary, inhomogeneous TDSE. Sequential fragmentation can still
occur, only that we cannot resolve it separately from non-sequential
fragmentation. We may also make the same statement in reverse: the
cost incurred by resolving sequential fragmentation is the need to
account for the transfer term \eqref{eqn:transfer-term}.


In addition to these interesting questions, the transfer terms have
practical impact on our numerical work. For the EOMs
\eqref{eqn:fragmentation-channel-eoms-matrix-form} to be solvable, we
require that the fragmentation channel overlap matrix \(\mat{B}\) is
non-defective, such that it may be diagonalized (at least formally),
or, as is implied in the time-stepping scheme
\eqref{eqn:time-stepping-scheme}, inverted. As long as the matching
boundary \(\domainboundary\) is chosen far enough away from the local
support of the ionic states \(\{\ket{I}\}\), \(\mat{B}\) reduces as
noted previously to the identity matrix, and the problem does not
occur. As the boundary is moved inwards, the fragmentation overlaps
\(\matrixel{I\vec{v}}{\fragmentationprojector}{I\vec{v}}\) grow, and
at some point the norm of the off-diagonal blocks may have increased
enough as to render \(\mat{B}\) defective. This situation may occur if
the ionic state \(\ket{I}\) is chosen as a sufficiently highly excited
Rydberg state, whose radial extent can be substantial. The practical
remedy is simple: increase the domain \(\domain\) enough to make
\(\mat{B}\) non-defective; for ease of implementation it is even
preferable to increase \(\domain\) such that
\(\mat{B}=\identity\). This would also improve the assumption that the
Hamiltonian in \(\complementdomain\) is negligibly different from the
scattering one: \(\Hamiltonian \approx \scatteringHamiltonian\).


\subsubsection{Infinite Time: iSURF}
\label{eqn:isurf}

When the laser field is over, the Hamiltonian is no longer time
dependent, in that the ion channel coupling matrix
\eqref{eqn:ion-channel-coupling-matrix} vanishes. The EOMs
\eqref{eqn:fragmentation-channel-eoms-matrix-form} then simplify to
give
\begin{equation}
  \label{eqn:fragmentation-channel-eoms-no-laser}
  \partial_t(\mat{B}\vec{a}) =
  \im\vec{S}.
\end{equation}
We may then compute an infinite-time correction to the amplitudes by
formally integrating \eqref{eqn:fragmentation-channel-eoms-no-laser}
to \(t=+\infty\). This forms the basis of the iSURF method
\cite{Morales2016-isurf}; the integral is performed by computing the
action of the resolvent on the final wavefunction
\(\ket{R(E)} \defd (\Hamiltonian - E)^{-1}\ket{\Psi(t_f)}\), for each
target energy of interest, and forming the source term from the
auxiliary state \(\ket{R(E)}\). Subsequently, we transform the
fragmentation amplitudes at the final time \(t_f\) as
\(\mat{B}(t_f)\vec{a}(t_f)\), and add the correction
\(\mat{B}(+\infty)\vec{a}(+\infty)-\mat{B}(t_f)\vec{a}(t_f)\) computed from
\eqref{eqn:fragmentation-channel-eoms-no-laser}. However, to avoid
these complications arising from the overlap matrix, as mentioned in
the preceding section \ref{sec:transfer-term-interpretation}, it is
advisable to increase \(\domain\) such that
\(\mat{B}=\identity\). Then the correction simplifies to
\(\vec{a}(+\infty)-\vec{a}(t_f)\).


\subsubsection{Example: Single and Double Ionization of \ch{He}}
\label{sec:fragmentation-channels-He-EOMs}
We now return to our earlier example in
section~\ref{sec:fragmentation-channels-He} of single and double
ionization of helium, and write down the amplitude EOMs explicitly,
which is possible due to the relative simplicity of this system. We
compare this with the approaches presented by
\cite{Scrinzi2012NJoP,Selstoe2021}. For moderate strengths of the
external field, we as stated previously expect the following channels:
\ch{He^*} (i.e.\ a possibly excited neutral atom), \ch{He+ + e-}, and
\ch{He^{2+} + 2 e-}. Assuming we are purely interested in
momentum-resolving the electronic degrees of freedom, we are left with
the asymptotic decomposition
\begin{equation}
  \fragmentationprojector\ket{\Psi(t)} =
  a_{I\vec{k}}(t)
  \ket{I\vec{k}} +
  a_{\vec{k}_1\vec{k}_2}(t)
  \ket{\vec{k}_1\vec{k}_2},
\end{equation}
where for simplicity we have suppressed the spin state. The vector of
source terms will have one entry for each ionic states \(\ket{I}\),
\(\ket{J}\), ..., corresponding to different single-ionization
channels, and one final entry for the double-ionization channel:
\begin{equation}
  \vec{S} =
  \bmat{
    \matrixel{\vec{k}}{\comm{\scatteringhamiltonian(t)}{\heaviside}}{\tilde{I}} \\
    \matrixel{\vec{k}}{\comm{\scatteringhamiltonian(t)}{\heaviside}}{\tilde{J}} \\
    \vdots \\
    \matrixel{\vec{k}_1\vec{k}_2}{\comm{\scatteringhamiltonian(t)}{\complementmapping}}{\Psi}
  } =
  \bmat{\vec{S}_s \\ \vec{S}_d}.
\end{equation}
Similarly, the fragmentation channel overlap matrix will be the
identity matrix, augmented in the last row and column by the
time-integrated transfer terms
\eqref{eqn:fragmentation-channel-overlap-matrix}:
\begin{equation}
  \label{eqn:He-fragmentation-channel-overlap-matrix}
  \mat{B} =
  \bmat{
    1 & 0 & \cdots & B_{I\vec{k}}^{\vec{k}_1\vec{k}_2} \\[1ex]
    0 & 1 & \cdots & B_{J\vec{k}}^{\vec{k}_1\vec{k}_2} \\[1ex]
    \vdots & \vdots & \ddots & \vdots \\[1ex]
    B_{\vec{k}_1\vec{k}_2}^{I\vec{k}} &
    B_{\vec{k}_1\vec{k}_2}^{J\vec{k}} & \cdots & 1 \\[1ex]
  } =
  \bmat{\identity_s & \mat{b} \\ \adjoint{\mat{b}} & \identity_d}
\end{equation}
\begin{equation}
  \mat{b} \in \Complex^{s\times d},
  \quad
  s = \#(I)\times\#(\vec{k}),
  \quad
  d = \#(\vec{k}_1\vec{k}_2).
\end{equation}
We note, that had we used exact scattering states of the field-free
helium Hamiltonian, \(\ket{I\vec{k}}\) and
\(\ket{\vec{k}_1\vec{k}_2}\) would be strictly orthogonal, and
\eqref{eqn:He-fragmentation-channel-overlap-matrix} would reduce to
the identity matrix. Finally, the ion channel coupling matrix only
couples single-ionization fragmentation channels, i.e.\ the last row
and column vanish:
\begin{equation}
  \tilde{\mat{H}} =
  \bmat{
    0 & \tilde{H}_{IJ} & \cdots & 0 \\[1ex]
    \tilde{H}_{JI} & 0 & \cdots & 0 \\[1ex]
    \vdots & \vdots & \ddots & \vdots \\[1ex]
    0 &
    0 & \cdots & 0 \\[1ex]
  }.
\end{equation}

At this point, we may compare the present work with that of
\citet{Scrinzi2012NJoP}; he describes the analysis of single and
double ionization of a two-electron system. Our approach is naturally
inspired by this particular work. Similarly to us, he devices a
two-stage approach, which separates single and double ionization into
separate classes. At variance with our approach, however, the first
ionization stage is described using a product state consisting of a
photoelectron state (a Volkov wave) and a ionic state with a
non-trivial time dependence, found by back-propagating the asymptotic
field-free ionic eigenstates from the detection time \(T\) to the
interaction time \(t\). In contrast, we directly accumulate the
amplitudes associated with the final ionic states through the use of
Dyson orbitals. The next, and more interesting difference, is the fact
that only \enquote{sequential} ionization is considered, in that
double ionization is accounted through the single ionization pathway;
only after the first electron has left the inner region, may the
second electron do the same. Indistinguishability of the electrons is
enforced by explicit permutation-symmetrization of the total
wavefunction. In our parlance, this approach is equivalent to having
the source term \(S_{I\vec{v}}(t)\) present in the EOMs
\eqref{eqn:fragmentation-channel-eoms-matrix-form} for the
single-ionization amplitude only, and double ionization is accounted
for using a transfer term \(T_{I\vec{v}}^{I'\vec{v}'}(t)\) \footnote{In
  figure~7 of \cite{Scrinzi2012NJoP}, a scheme including also the
  source term for non-sequential double ionization is proposed, but
  not discussed any further.}. Finally, the conjugate transfer terms
are not included, which precludes any recombination.

\subsection{Analytic Fragment Wavefunctions}
\label{sec:analytic-fragment-wavefunctions}
In this section, we discuss various analytic approximations to the
fragment wavefunctions \(\ket{\phi_i}\).

\subsubsection{Volkov Waves}
\label{sec:volkov-waves}
These are eigenstates of the Volkov Hamiltonian
\eqref{eqn:schroedinger-one-body-scattering-hamiltonian}. For
non-relativistic theories, they only attain a simple form in the
dipole approximation \cite{He2017}. Assuming the dipole approximation
below, in the velocity gauge, the Volkov waves are given by
\begin{equation}
  \chi_{\vec{p}v}^V(t) =
  \exp[-\im\Lambda(\vec{p},t)]
  \chi_{\vec{p}},
\end{equation}
where
\begin{equation}
  \chi_{\vec{p}} \defd
  \frac{\ce^{\im\vec{p}\cdot\vec{r}}}{(2\pi)^{n/2}}
\end{equation}
is the unperturbed plane wave component, and
\begin{equation}
  \Lambda(\vec{p},t) \defd
  \int_{t_{\textrm{ref}}}^t\diff{t'}
  \frac{[\vec{p}-q\vec{A}(t')]^2}{2m}
\end{equation}
is the \emph{Volkov phase} [i.e.\ the time-integrated Volkov
Hamiltonian
\eqref{eqn:schroedinger-one-body-scattering-hamiltonian}]. \(t_{\textrm{ref}}\)
is some reference time, chosen at our convenience.

We find the length gauge form by employing the gauge transform
\cite{Bandrauk2013JoPBAMaOP}
\begin{equation}
  \label{eqn:volkov-wave-length-gauge}
  \begin{aligned}
    \chi_{\vec{p}l}^V(t)
    &=
      \exp[-\im q\vec{A}(t)\cdot\vec{r}]
      \chi_{\vec{p}v}^V(t) \\
    &=
      \exp[-\im\Lambda(\vec{p},t)]
      \frac{\ce^{\im\vec{k}(t)\cdot\vec{r}}}{(2\pi)^{n/2}}, \\
    \vec{k}(t)
    &\defd
      \vec{p}-q\vec{A}(t),
  \end{aligned}
\end{equation}
i.e.\ the plane wave component is no longer stationary, but oscillates
with the laser field.

As was already alluded to in section~\ref{sec:cartesian-coordinates},
employing Volkov waves as the scattering states in
\eqref{eqn:surface-term-current} amounts to \((N-1)\)-dimensional
Fourier transforms of the amplitude and derivative of the Dyson
orbital at the boundary \(\domainboundary\). This is not true only in
Cartesian coordinates, but in any geometry; in e.g.\ spherical
coordinates we recognize the Fourier transform in the spherical Bessel
functions employed there
\cite{Tao2012NJoP,Morales2016-isurf}. However, the spatial derivatives
of the Volkov waves are particularly simple in Cartesian coordinates:
\begin{equation}
  \label{eqn:volkov-spatial-derivative}
  \begin{aligned}
    \partial_u^q
    \chi_{\vec{p}v}^V(t)
    &=
      (\im p_u)^q
      \chi_{\vec{p}v}^V(t), \\
    \partial_u^q
    \chi_{\vec{p}l}^V(t)
    &=
      [\im k_u(t)]^q
      \chi_{\vec{p}l}^V(t).
  \end{aligned}
\end{equation}

\subsubsection{Many-Body Volkov Waves}


The lowest-order approximation is the Volkov superwave
\(\exp(\im kr)\), where
\begin{equation}
  \begin{aligned}
    k
    &=
      \bmat{\vec{k}_1 \\
    \vec{k}_2 \\
    \vdots \\
    \vec{k}_v} =
    \bmat{\vec{p}_1-q_1\vec{A}(\vec{r}_1,t) \\
    \vec{p}_2-q_2\vec{A}(\vec{r}_2,t) \\
    \vdots \\
    \vec{p}_v-q_v\vec{A}(\vec{r}_v,t)},
    &
      r
    &=
      \bmat{\vec{r}_1 \\
    \vec{r}_2 \\
    \vdots \\
    \vec{r}_v}.
  \end{aligned}
\end{equation}
We must be careful not to forget the mass when applying the scattering
Hamiltonian \eqref{eqn:schroedinger-one-body-scattering-hamiltonian},
since it may differ between different fragments.

\subsubsection{Coulomb--Volkov Waves; Coulomb Waves}

There is no known analytic solution to the Coulomb--Volkov
problem. Analytic approximations such as the eikonal Volkov
approximation \cite{Smirnova2008}, while very successful in
strong-field approximation calculations, are less applicable to
numerical calculations, due to being only approximately orthogonal.

If we instead find the exact Coulomb--Volkov waves numerically, we
have to solve an auxiliary problem of single ionization of a
hydrogen-like system numerically, \emph{for every \(\vec{k}\) of
  interest}, completely eradicating any advantage of the tSURFF
approach which is based on the efficient evaluation of approximate
scattering states over a range of momenta. This approach also scales
badly with respect to number of ionized electrons.

The situation improves somewhat if we can afford to keep the full
wavefunction in the computational box, i.e.\ all fragments remain
within \(\domain\) after all external fields have vanished. We may
then employ the field-free Coulomb waves in conjunction with the iSURF
technique \cite{Morales2016-isurf}, as described in
section~\ref{eqn:isurf}. This is useful when accurate computation of
low-energy fragments is desired.

\subsection{Comparison with \(R\)-Matrix Theory}
\label{sec:R-matrix-comparison}

A very similar problem is encountered in \(R\)-matrix theory
\cite{Burke2005}, where again, the Hilbert space \(\HilbertSpace\) is
partitioned into an inner (\(+\)) and an outer (\(-\)) region:
\(\HilbertSpace = \Pspace_+ \cup \Pspace_-\). The TDSE then partitions
as
\begin{equation}
  [\imdt - \blochHamiltonian_\pm(t)]
  \ket{\Psi_\pm(t)} =
  \bloch_\mp
  \ket{\Psi_\mp(t)},
\end{equation}
where \(\blochHamiltonian_\pm(t) \defd \Hamiltonian(t) + \bloch_\pm\),
and the \emph{Bloch operator}
\(\bloch_\pm \defd -\comm{\proj_\pm}{\Hamiltonian}\) connects the
wavefunction across the boundary between the regions. We may then
easily show (see appendix~\ref{app:R-matrix-comparison}) that the
fragmentation amplitude can be written as
\begin{equation}
  \label{eqn:R-matrix-fragmentation-amplitude}
  a_{I\vec{v}}(T) =
  -\im\int_{t_i}^T\diff{t}
  \propU_{I\vec{v}I'\vec{v}}(T,t)
  \matrixel{I'\vec{v}(t)}{\bloch_+}{\Psi(t)},
\end{equation}
where \(\propU_{I\vec{v}I'\vec{v}}(T,t)\) is the propagator associated
with the fragmentation channel \(\vec{v}\). The matrix element
\(\matrixel{I'\vec{v}(t)}{\bloch_+}{\Psi(t)}\) is of course directly
related to our source term \(S_{I'\vec{v}}(t)\), and by introducing an
expansion such as \eqref{eqn:asymptotic-ansatz}, we will also recover
the transfer terms \(T_{I\vec{v}}^{I'\vec{v}'}(t)\).

Similarly, the time-independent Schrödinger equation
\([\Hamiltonian-E]\ket{\Psi} = 0\) also breaks up into subspaces as
\cite[cf. Eqs.~(8) and (32) of ][]{Burke2005}
\begin{equation}
  \label{eqn:partitioned-tise}
  [\blochHamiltonian_\pm-E]
  \ket{\Psi_\pm} =
  -\bloch_\mp
  \ket{\Psi_\mp},
\end{equation}
the solution of which is given by \cite[cf. Eqs.~(9) and (34) of
][]{Burke2005}
\begin{equation}
  \label{eqn:partitioned-tise-solution}
  \ket{\Psi_\pm} =
  -[\blochHamiltonian_\pm-E]^{-1}
  \bloch_\mp
  \ket{\Psi_\mp}.
\end{equation}

The main philosophical differences between \(R\)-matrix theory and the
surface-flux methods are twofold: Firstly, in the outer region,
\(R\)-matrix methods typically solve the Schrödinger equation
numerically exactly, under the assumption that exchange interaction
can be neglected, whilst still keeping a long-range, multipolar
potential accounting for the charge distribution of the nuclei and
electrons in the inner region. Surface-flux methods additionally
neglect the long-range potentials in the outer region, and the
Schrödinger equation in that region is solved approximately, but
analytically.

Secondly, the matching of solutions in the inner and outer regions in
\(R\)-matrix theory is accomplished by the eponymous \(R\) matrices,
which are in essence spectral decompositions of the resolvent operator
\((E-\Hamiltonian)^{-1}\), applied to the inner region solution. This
provides boundary conditions for the solution of the scattering
problem \eqref{eqn:partitioned-tise-solution} in the outer region, for
any desired total energy \(E\) \cite[for details on the procedure,
see][]{Burke2005}. In the surface-flux methods, we, as we have seen
above in section~\ref{sec:probability-in-complement-domain}, instead
compute a surface integral between the wavefunction and the analytic
scattering state. We thereby forego the spectral decomposition of the
former, which can potentially be a costly procedure.

\section{Simple Demonstration}
\label{sec:demonstration}

As a simple demonstration of multiple fragmentation, we employ two
one-dimensional, multipartite systems. For both systems, the TDSE
reads
\begin{equation}
  \imdt\ket{\Psi} =
  \Hamiltonian(t)\ket{\Psi},
\end{equation}
\begin{equation}
  \Hamiltonian(t)
  \defd
  \begin{cases}
    \displaystyle
    \frac{\operator{p}_i^2}{2m_i} +
    \frac{1}{2}\potop_{ij} -
    q_i\operator{x}_iF(t),
    &
      \textrm{(length gauge)}, \\[1em]
    \displaystyle
    \frac{[\operator{p}_i - q_iA(t)]^2}{2m_i} +
    \frac{1}{2}\potop_{ij},
    &
      \textrm{(velocity gauge)},
  \end{cases}
\end{equation}
where \(\operator{p}_i\) is the momentum of the \(i\)th particle, with
charge \(q_i\) and mass \(m_i\), \(A\) is the vector potential (which
we for simplicity assume to be spatially independent), and
\(F(t)\defd-\partial_t A(t)\) is the electric field. Together, they
form the the kinetic operator,
\begin{equation}
  \kinop(t)
  \defd
  \begin{cases}
    \displaystyle
    \frac{\operator{p}_i^2}{2m_i},
    &
      \textrm{(length gauge)}, \\[1em]
    \displaystyle
    \frac{[\operator{p}_i - q_iA(t)]^2}{2m_i},
    &
      \textrm{(velocity gauge)},
  \end{cases}
\end{equation}
which, along with is exponential \(\ce^{-\im\kinop(t)}\), is most
naturally evaluated in Fourier space. \(\potop_{ij}\) is the
potential between the particles \(i\) and \(j\), only depending on
their relative positions:
\begin{equation}
  \label{eqn:soft-core-potential}
  \potop_{ij} \defd
  q_iq_jV(x_i - x_j), \quad
  V(r) \defd
  \frac{\ce^{-\alpha r}}{\sqrt{r^2 + \varepsilon}}, \quad
  \alpha \ge 0, \quad \varepsilon > 0;
\end{equation}
it is attractive for \(q_iq_j < 0\) and repulsive for
\(q_iq_j>0\). The smoothening parameter \(\varepsilon\) ensures the potential is
never singular, whereas the optional screening parameter \(\alpha\), if
non-zero, limits the range of the potential.

Computing derivatives and integrals in (truncated) Fourier space is
commensurate with periodic boundary conditions. To avoid having a
wavepacket that moves through the boundary on the right side of the
domain reappear on the left side, all operations involving Fourier
transforms occur on a grid twice the size of the computational
domain. Additionally, a complex-absorbing potential
\cite{Manolopoulos2002} is employed in real space, to limit the amount
of reflections.

One step with the time propagator is thus approximated as
\begin{equation}
  \timeord\exp\left[-\im\int_0^h\diff{t}\Hamiltonian(t)\right]
  \approx
  \ce^{-\frac{\im h}{2}\potop(h/2)}
  \ce^{-\im h\kinop(h)}
  \ce^{-\frac{\im h}{2}\potop(h/2)},
\end{equation}
where \(\timeord\) is the time-ordering operator and
\begin{equation}
  \potop(t) =
  \operator{V}_{ij} +
  \potop_{\textrm{CAP}} -
  \begin{cases}
    q_i\operator{x}_iF(t),
    &
      \textrm{(length gauge)}, \\
    0,
    &
      \textrm{(velocity gauge)}.
  \end{cases}
\end{equation}

The scattering states are approximated using Volkov waves, as
described in section~\ref{sec:volkov-waves} (see also
section~\ref{sec:cartesian-coordinates} for a general discussion on
Cartesian coordinate systems).

The amplitudes and the derivatives of the Dyson orbitals at the
matching surfaces are computed through quintic Hermite polynomial
interpolation, where the requisite derivatives are approximated using
a fourth-order-accurate stencil found using the algorithm by
\textcite{Fornberg1988,Fornberg1998}.

\subsection{Single Ionization of 1d \ch{H}}
\label{sec:single-ionization-1d-H}

\begin{figure}[htb]
  \centering
  \includegraphics[trim=30cm 30cm 30cm 30cm]{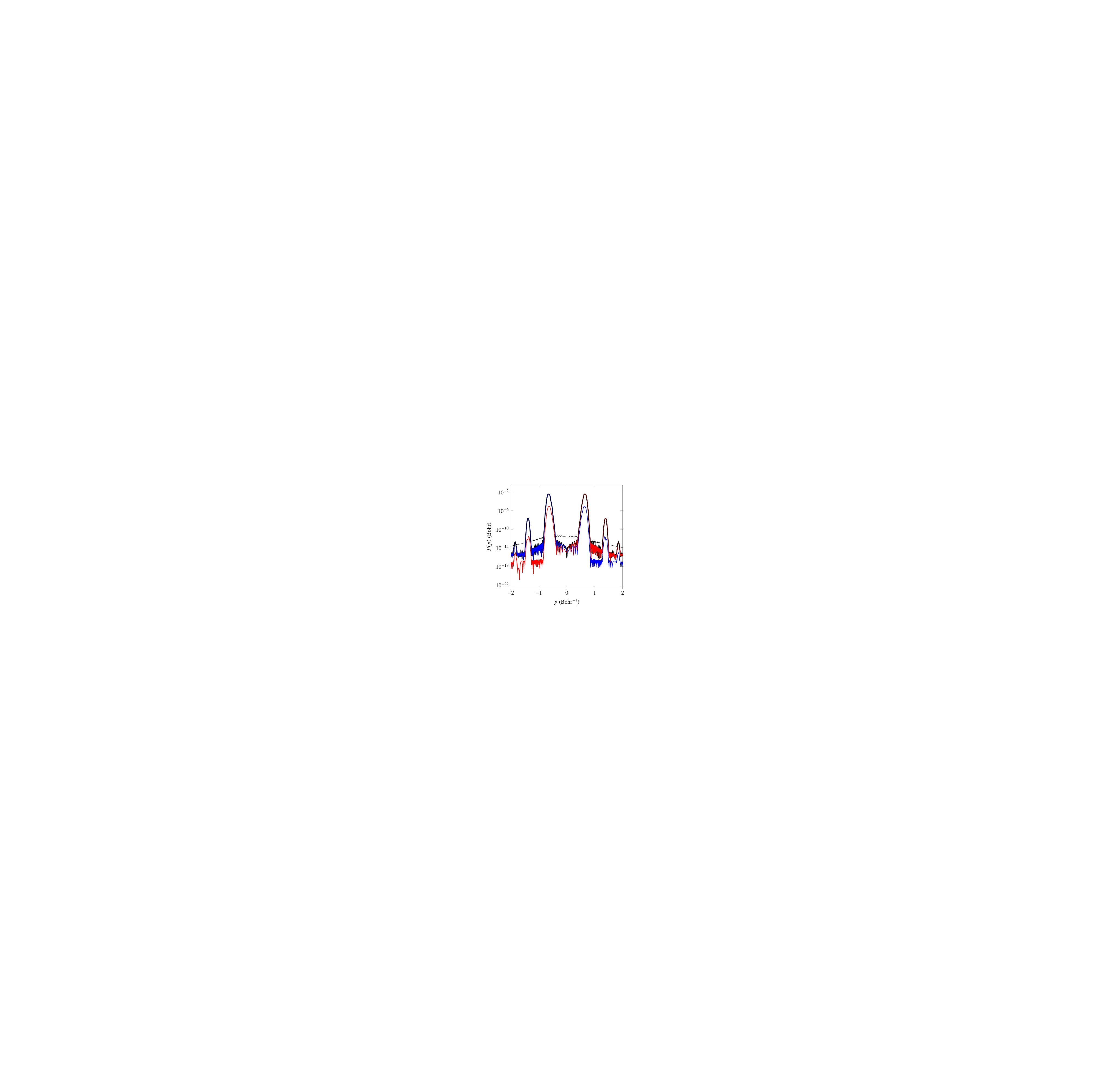}
  \caption{\label{fig:single-ionization-h} The thick black line shows
    the total photoelectron spectrum; the blue line shows the
    contribution to the spectrum from the surface at
    \(x=\SI{-100}{Bohr}\), the red line shows the contribution to the
    spectrum from the surface at \(x=\SI{100}{Bohr}\), and the thin
    black line shows the reference spectrum computed using the FFT,
    after projecting out all the bound states from the final
    wavefunction (it overlaps the peaks almost perfectly, only the
    noise floor is higher).}
\end{figure}

\begin{figure*}
  \centering
  \includegraphics[trim=30cm 30cm 30cm 30cm]{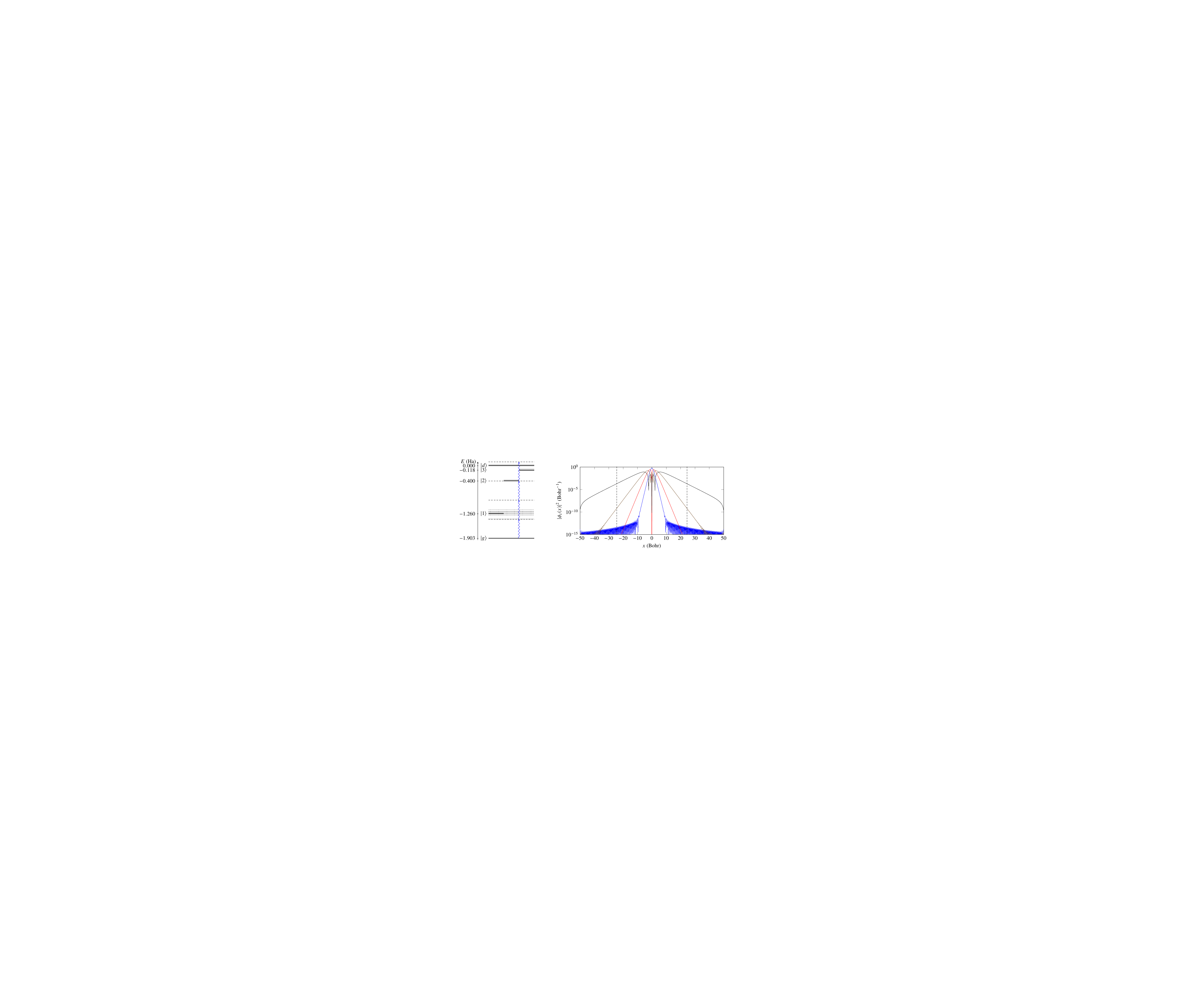}
  \caption{\emph{Left}: \ch{He} energy level scheme; \(\ket{g}\)
    denotes the ground state, \(\ket{n}\), \(n=1,2,3\) single
    ionization thresholds associated with states of \ch{H}, and
    \(\ket{d}\) the double ionization threshold. The dotted lines
    indicate the positions of the first few excited two-electron
    states, which may contributed to resonant-enhanced multiphoton
    ionization. The dashed, horizontal lines and the wavy arrows
    indicate expected photoionization peaks after absorption of one
    and two photons of energy \(\hbar\omega=1/2\), respectively. \emph{Right}:
    The first four states of \ch{He+} (corresponding to
    \(\ket{g},\ket{n}\) on the left), as a function of the
    one-electron coordinate \(x\). The dashed, vertical lines indicate
    the boundary \(\Sphere^0(r)\) with \(r\approx\SI{24.4}{Bohr}\).}
  \label{fig:he-energy-level-scheme}
\end{figure*}

This first, trivial, example investigates single ionization of 1d
\ch{H}. The ground state of the cation is the vacuum, and thus, the
wavefunction \emph{is} the Dyson orbital. The ionizing field consists
of \SI{1}{\tera\watt\per\centi\meter\squared} at
\(\hbar\omega=\SI{0.75}{\hartree}\). The envelope is a Gaussian of
\SI{1.45}{\femto\second} duration, and it is truncated at \(6\sigma\). The
calculation continues for \SI{10}{\femto\second} after the end of the
pulse, to allow the photoelectron wavepacket time to pass through the
boundary \(\domainboundary\).

We employ a uniform grid extending from \SIrange{-50}{50}{Bohr}, with
a grid spacing of \(\Delta x\approx\SI{0.1}{Bohr}\). The CAP by
\citet{Manolopoulos2002} is designed to reflect
\(\le\SI{1}{\percent}\) for kinetic energies \(\ge\SI{0.1}{Ha}\), which
leads to a CAP encompassing the last \SI{14.05}{Bohr} of the
computational box in all directions. With a safety margin of
\(\SI{10}{Bohr}+2\alpha\), where
\(\alpha\defd F/\omega^2\approx\SI{9.5}{\milli{Bohr}}\) is the free oscillation
amplitude of the driving field, we place the boundary at
\(r = \SI{50}{Bohr}-\SI{14.05}{Bohr} - \SI{10}{Bohr} - 2\alpha \approx
\SI{25.93}{Bohr}\).

The binding potential \eqref{eqn:soft-core-potential} is used with a
smoothing parameter of \(\varepsilon = 1\) and a screening parameter of
\(\alpha = 0.3\), which yields a ground state energy of
\SI{-0.538}{\hartree}. The resulting photoelectron spectrum is shown
in figure~\ref{fig:single-ionization-h}. Excellent agreement is
achieved with the reference spectrum computed via FFT from a
wavefunction resolved on a larger grid extending to
\(\pm\SI{400}{Bohr}\).

\subsection{Single and Double Ionization of 1d \ch{He}}

\begin{figure}[htb]
  \centering
  \includegraphics[trim=30cm 30cm 30cm 30cm]{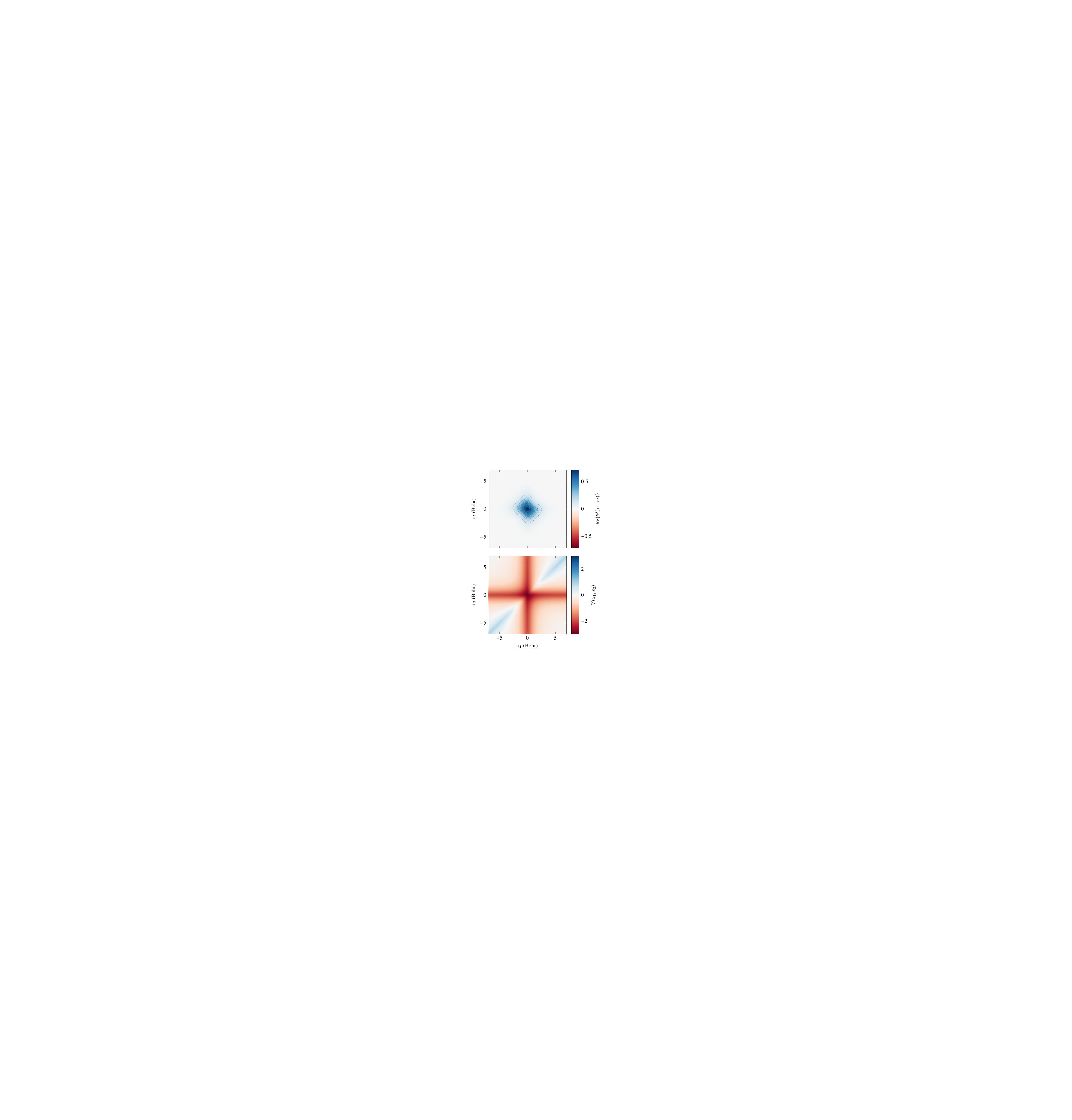}
  \caption{\ch{He} two-electron ground state (\emph{top}) and
    potential (\emph{bottom}). Dashed contours are isolines of
    \(\Re\{\Psi(x_1,x_2)\}\) and solid contours are isolines of
    \(\abs{\Psi(x_1,x_2)}^2\).}
  \label{fig:he-two-electron-state}
\end{figure}

\begin{figure}[htb]
  \centering
  \includegraphics[trim=30cm 30cm 30cm 30cm]{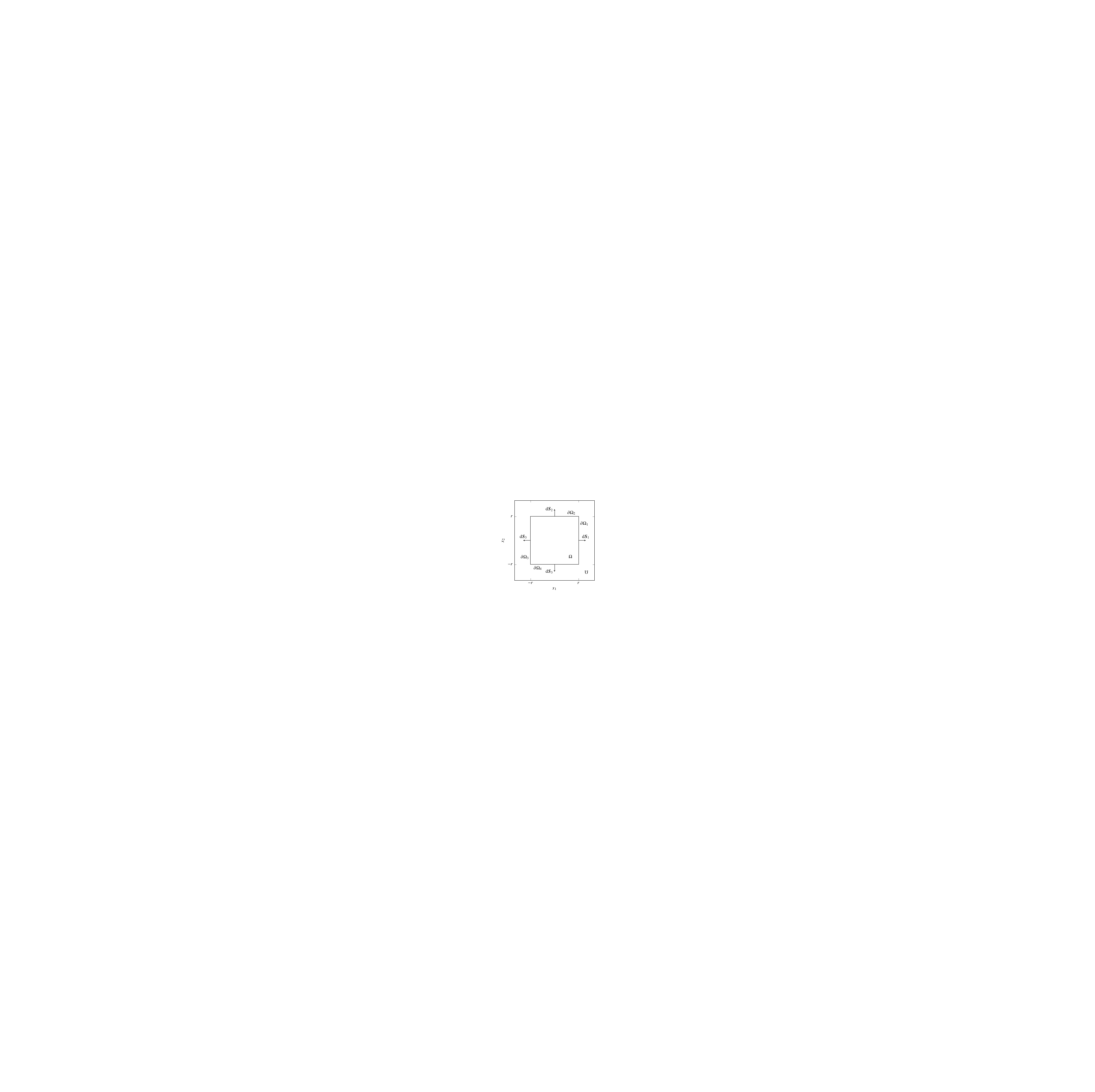}
  \caption{Double ionization geometry. For ease of reference, we have
    labelled the piecewise boundary as \(\domainboundary_i\),
    \(i=1,2,3,4\), and indicated their respective surface integration
    measures \(\diff{\vec{S}_i}\).}
  \label{fig:he-double-ionization-geometry}
\end{figure}

The next example is 1d \ch{He}, which may undergo both single and
double ionization. We assume the electrons are in a singlet state:
\(\frac{1}{\sqrt{2}}(\ket{\spinup\spindown} -
\ket{\spindown\spinup})\), and in the absence of spin--orbit coupling,
the spin-singlet symmetry is not broken. Since the electron spin
function is antisymmetric, the spatial function must be symmetric. For
simplicity, we assume that the nucleus is spin-less. This is the same
problem as treated by \citet{Grobe1992,Grobe1993,Grobe1993a}. We again
choose \(\varepsilon=1\) and \(\alpha=0.3\) in \eqref{eqn:soft-core-potential}, which
leads to the energy level scheme depicted in
figure~\ref{fig:he-energy-level-scheme}.

For single ionization, the source term
\eqref{eqn:source-term-divergence-form} collapses to the case of the
line \eqref{eqn:line-surface-integrals}:
\begin{equation}
  \begin{aligned}
    S_{Ip}(t) =
    \frac{1}{2}
    \frac{\exp[\im\Lambda(p,t)]}{\sqrt{2\pi}}
    &\int\diff{x}
      \ce^{-\im k(t)x}
      [-\im k(t)\tilde{I}(x) - \tilde{I}_{,x}(x)] \\
    &
    \hphantom{\int\diff{x}}\times
    [\delta(x-r) - \delta(x+r)] \\
    =
    \frac{1}{2}
    \frac{\exp[\im\Lambda(p,t)]}{\sqrt{2\pi}}
    &\{\ce^{-\im k(t)r}
      [-\im k(t)\tilde{I}(r) - \tilde{I}_{,x}(r)] \\
    &\hphantom{\{}-
      \ce^{\im k(t)r}
      [-\im k(t)\tilde{I}(-r) - \tilde{I}_{,x}(-r)]\},
  \end{aligned}
\end{equation}
here shown in the length gauge. As mentioned earlier, \(\tilde{I}(r)\)
is the Dyson orbital, and \(\tilde{I}_{,x}(r)\) its spatial
derivative, both evaluated at the matching boundary
\(\domainboundary\), \(x=r\). It is formed by projecting the
time-dependent wavefunction onto the ionic state as
\(\ket*{\tilde{I}(t)} = \ce^{\im E_I(t-t_i)}
\sqrt{2}\brakett{I(t_i)}{\Psi}\). The kinematic momentum is given by
\begin{equation}
  k(t) = p + A(t),
\end{equation}
since electrons have charge \(q=-\abs{e}\).

For double ionization, the geometry is as sketched in
figure~\ref{fig:he-double-ionization-geometry}. In this case, we have
no ionic state, but directly project the full wavefunction onto the
two-electron continuum, which we approximate using the Volkov
superwave of two electrons with canonical momenta \(p_1\) and \(p_2\),
respectively. Again in the length gauge, we thus have
\begin{equation}
  \begin{aligned}
    S_{\vec{p}}
    &=
      \frac{1}{2}
      \oint_{\domainboundary}
      \!\!\!\diff{S}\;
      g^{uv}
      [\tensor{V}{^*_{,u}}
      \Psi -
      \conj{V}
      \tensor{\Psi}{_{,u}}]
      n_v \\
    &=
      \frac{1}{2}
      \left(
      \oint_{\Pi_j}
      \!\!\!\diff{S}
      -
      \oint_{\Pi_{-j}}
      \!\!\!\!\!\diff{S}
      \right)
      [\tensor{V}{^{*,j}}
      \Psi -
      \conj{V}
      \tensor{\Psi}{^{,j}}] \\
    &=
      \frac{1}{2}
      \operator{\varsigma}_j
      \chi_{\vec{p}l}^{V*}(t)
      [-\im k_j(t) -
      \partial_j]
      \Psi \\
    &=
      \frac{1}{2}
      \frac{\exp[\im\Lambda(\vec{p},t)]}{2\pi}
      \operator{\varsigma}_j
      \ce^{-\im\vec{k}(t)\cdot\vec{r}}
      [-\im k_j(t) -
      \partial_j]
      \Psi
  \end{aligned}
\end{equation}
where \(j=1,2\), \(\vec{r} \defd \bmat{x_1 \\ x_2}\),
\(\vec{p} \defd \bmat{p_1 \\ p_2}\),
\begin{equation*}
  \vec{k}(t)
  \defd
  \bmat{k_1(t) \\ k_2(t)} =
  \bmat{p_1 - q_1 A(t) \\ p_2 - q_2 A(t)} \equiv
  \bmat{p_1 + A(t) \\ p_2 + A(t)},
\end{equation*}
\begin{equation}
  \operator{\varsigma}_jU^j
  \defd
  \oint_{\Pi_j}
  \!\!\!\diff{S}\;
  U^j
  -
  \oint_{\Pi_{-j}}
  \!\!\!\!\!\diff{S}\;
  U^j,
\end{equation}
and we have inserted the expression
\eqref{eqn:volkov-wave-length-gauge} for the Volkov waves in the
length gauge. As mentioned in \ref{sec:cartesian-coordinates}, the
surface integrals are most efficiently evaluated using FFTs; when
evaluating the surface integral over the line crossing \(x_i\), the
FFT over the other coordinate \(x_j\) immediately yields the sought
overlap for all momenta \(k_j\). By padding the extracted surface
values and derivatives with zeros before employing the FFT, we
efficiently interpolate onto a finer grid in \(k\), enabling us to
resolve finer features such as \emph{Freeman resonances}
\cite{Freeman1987}.

The ionization is driven by \SI{e15}{\watt\per\centi\meter\squared}
radiation at \(\hbar\omega=\SI{0.5}{Ha}\), with a Gaussian envelope of
\SI{1.45}{\femto\second}, again truncated at \(6\sigma\), and additional
post-propagation of \SI{10}{\femto\second}.

The grid parameters are the same as employed in the previous example,
see \ref{sec:single-ionization-1d-H}, except for the use of a slightly
coarser grid with \(\Delta x = \SI{0.4}{Bohr}\). Given the larger free
oscillation amplitude of the driving field of
\(\alpha\approx\SI{0.68}{Bohr}\), we place the boundary
\(\domainboundary\) at \(r = \SI{24.40}{Bohr}\).

\begin{figure}
  \includegraphics[trim=30cm 30cm 30cm 30cm]{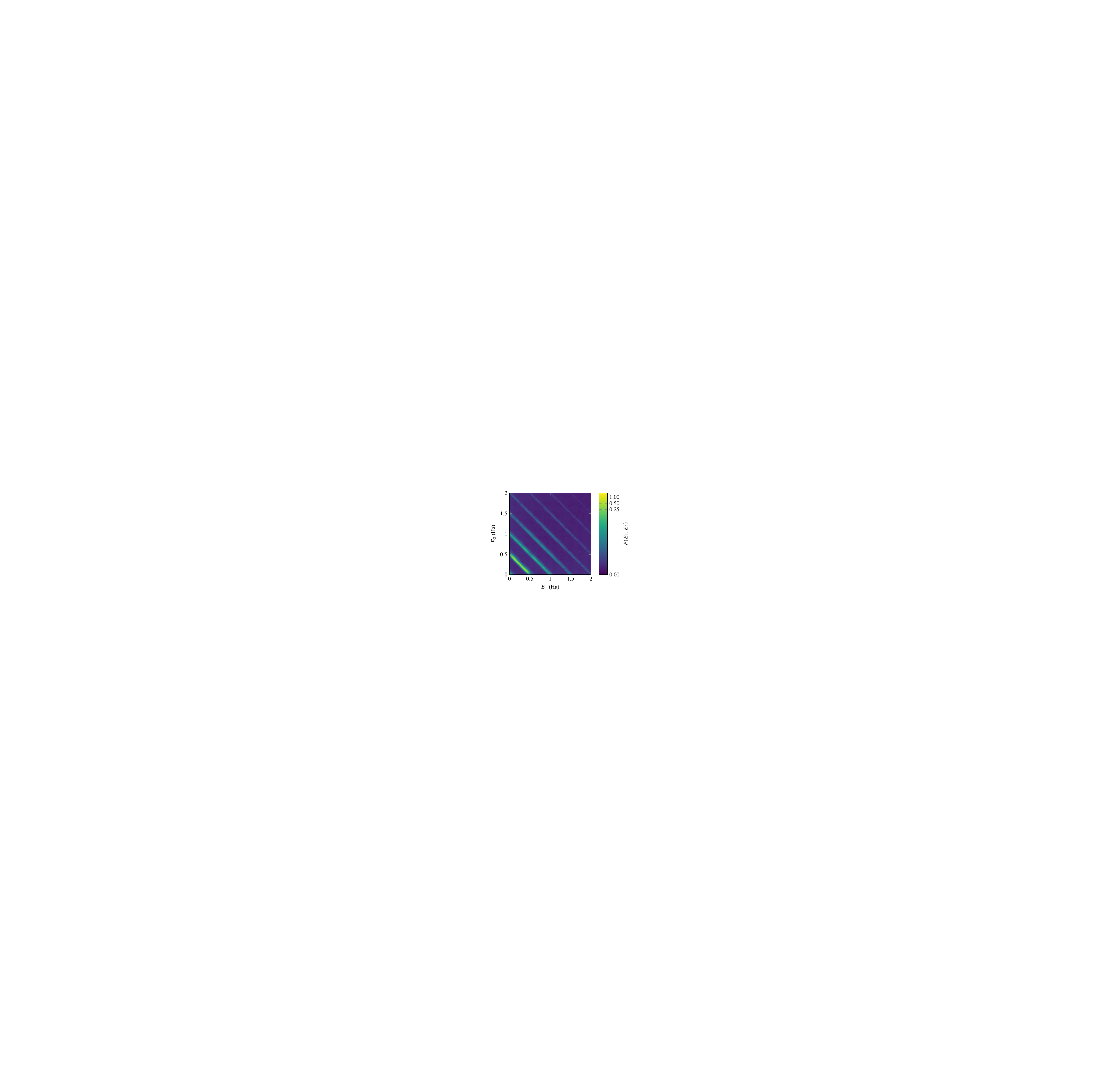}
  \caption{\label{fig:double-ionization-he} Double ionization of
    \ch{He}. The diagonal lines are the hallmark of
    \emph{energy-sharing} photoelectron peaks
    \cite{Pazourek2011,Scrinzi2012NJoP}, i.e.\
    \(E_1+E_2 = q\hbar\omega - I_p\), where \(q\) is the number of photons
    absorbed, and \(I_p\) is the double ionization threshold.}
\end{figure}

\begin{figure*}
  \centering
  \includegraphics[trim=30cm 30cm 30cm 30cm]{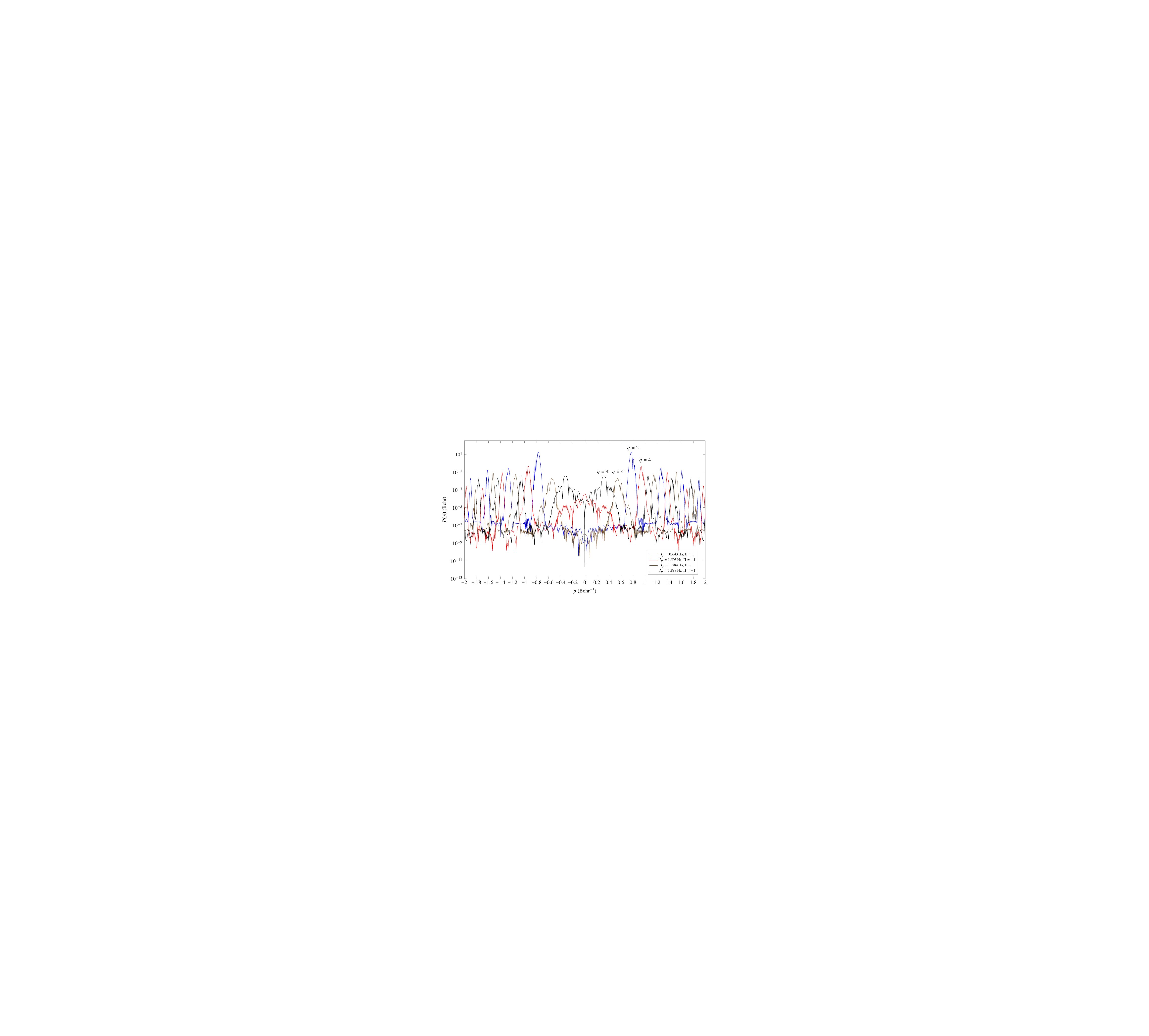}
  \caption{\label{fig:single-ionization-he} Single ionization of
    \ch{He}, resolved on the ionic states of \ch{He+}. Indicated above
    the first peak of each ionization channel is the number of photons
    necessary to reach that particular peak;
    cf. figure~\ref{fig:he-energy-level-scheme}.
  }
\end{figure*}

The resulting single-ionization photoelectron spectrum, resolved on
eigenstates of the cation \ch{He+}, is shown in
figure~\ref{fig:single-ionization-he}. Clearly visible superposed on
each photoelectron peak are the Freeman resonances, due to
intermediate excitation of the two-electron excited states, as shown
in figure~\ref{fig:he-energy-level-scheme}. These are also visible in
the double-ionization photoelectron spectrum, shown in
figure~\ref{fig:double-ionization-he}.


\section{Discussion}
\label{sec:discussion}

We have derived a general framework for the systematic treatment of
multiple levels of fragmentation, fully differential on all degrees of
freedom desired, whilst still keeping the bookkeeping
manageable. Owing to the partition of space into an inner region and
an outer region, where the wavefunction can be decomposed into
asymptotic fragments, the computational and storage requirements can
be kept rather modest in comparison with other methods which require
the entire wavefunction to be kept in the computational box before the
analysis stage. We have also provided simple test examples to
demonstrate the differential power of the proposed method.


%
%

\begin{acknowledgments}
  SC would like to thank Johan Rathsman, Johan Bijnens, and Takeshi
  Sato for helpful discussions.
\end{acknowledgments}

\appendix

\section{Comparison with \(R\)-Matrix Theory}
\label{app:R-matrix-comparison}

Here, we present the derivation of the equations behind the
\(R\)-matrix theory, used for the comparison made in
section~\ref{sec:R-matrix-comparison} in the main text. The
partitioning of Hilbert space \(\HilbertSpace\) into an inner region
\(\Pspace_+\) and an outer region \(\Pspace_-\), can be accomplished
using two projectors:
\begin{equation}
  \proj_\pm\HilbertSpace =
  \Pspace_\pm.
\end{equation}
In our case, \(\proj_+\) and \(\proj_-\) are naturally given by
\eqref{eqn:mapping-operator} and
\eqref{eqn:complement-mapping-operator}, respectively. The projectors
fulfil the expected relations
\begin{equation}
  \begin{aligned}
    \proj_+ + \proj_-
    &=
      \identity,
    &
      \proj_m\proj_n
    &=
      \delta_{mn}
      \proj_m,
    &
      m,n
    &= +,-.
  \end{aligned}
\end{equation}
With the aid of these projectors, we may partition the Hamiltonian and
the wavefunction as
\begin{equation}
  \label{eqn:partitioned-hamiltonian}
  \Hamiltonian =
  \Hamiltonian_{++} +
  \Hamiltonian_{--} +
  \Hamiltonian_{+-} +
  \Hamiltonian_{-+},
\end{equation}
where
\begin{equation}
  \label{eqn:subspace-hamiltonians}
  \begin{aligned}
    \Hamiltonian_{mn}
    \defd
    \proj_m\Hamiltonian\proj_n
    &=
      \proj_m\proj_n\Hamiltonian -
      \proj_m\comm{\proj_n}{\Hamiltonian} \\
    &=
      \proj_m\left\{
      \delta_{mn}
      \Hamiltonian -
      \comm{\proj_n}{\Hamiltonian}
      \right\},
  \end{aligned}
\end{equation}
and
\begin{equation}
  \label{eqn:projected-wavefunction}
  \ket{\Psi} =
  \proj_m\ket{\Psi} \defd
  \ket{\Psi_m}.
\end{equation}
We can further rewrite \eqref{eqn:subspace-hamiltonians} as
\begin{equation}
  \tag{\ref{eqn:subspace-hamiltonians}*}
  \label{eqn:subspace-hamiltonians-with-bloch}
  \begin{aligned}
    \proj_\pm\Hamiltonian\proj_\pm
    &=
      \proj_\pm[\Hamiltonian + \bloch_\pm]
      \defd
      \proj_\pm\blochHamiltonian_\pm, \\
    \proj_\pm\Hamiltonian\proj_\mp
    &=
      \proj_\pm\bloch_\mp,
  \end{aligned}
\end{equation}
where the \emph{Bloch operator} \cite{Bloch1957} is defined as
\begin{equation}
  \label{eqn:bloch-operator}
  \bloch_\pm \defd
  -\comm{\proj_\pm}{\Hamiltonian}.
\end{equation}

The TDSE in subspace \(\Pspace_\pm\) is then reformulated as
\cite{Torlina2012PRA}
\begin{equation}
  \label{eqn:partitioned-tdse}
  [\imdt-\blochHamiltonian_\pm(t)]
  \ket{\Psi_\pm(t)} =
  \bloch_\mp
  \ket{\Psi_\mp(t)},
\end{equation}
where we gain an inhomogeneous term from the subspace
\(\Pspace_\mp\). Introducing the subspace evolution operators
\begin{equation}
  \label{eqn:subspace-propagators}
  \propU_\pm(a,b) \defd
  \timeord
  \exp\left[ -\im\int_b^a\diff{\tau} \Hamiltonian_{\pm\pm}(\tau) \right],
\end{equation}
where \(\timeord\) is the time-ordering operator, we can rewrite the
subspace TDSE \eqref{eqn:partitioned-tdse} on integral form,
reminiscent of the Dyson expansion \cite{Smirnova2007a}
\begin{equation}
  \tag{\ref{eqn:partitioned-tdse}*}
  \begin{aligned}
    \ket{\Psi_\pm(a)}
    &=
      \propU_\pm(a,b)
      \ket{\Psi_\pm(b)} \\
    &\hphantom{=}-
      \im
      \int_b^a\diff{t}
      \propU_\pm(a,t)
      \Hamiltonian_{\pm\mp}(t)
      \propU_\mp(t,b)
      \ket{\Psi_\mp(b)} \\
    &\equiv
      \propU_\pm(a,b)
      \ket{\Psi_\pm(b)} \\
    &\hphantom{=}-
      \im
      \int_b^a\diff{t}
      \propU_\pm(a,t)
      \bloch_\mp
      \propU_\mp(t,b)
      \ket{\Psi_\mp(b)}.
  \end{aligned}
\end{equation}
We may now introduce the \emph{retarded} and \emph{advanced Green's
  functions} \cite[see section~III.A.2 of][]{Cohen-Tannoudji1998}
according to
\begin{equation}
  \label{eqn:greens-functions}
  \greensfunction_\pm^\gamma(t,t')
  \defd
  \gamma
  \propU_\pm(t,t')
  \heaviside[\gamma(t-t')],
  \quad
  \gamma = \pm,
\end{equation}
where the subscript \(\pm\) as before refers to the inner (\(+\)) and
outer (\(-\)) regions, and the superscript \(\gamma=+\) refers to the
retarded (\(t>t'\)) Green's function, and \(\gamma=-\) to the advanced
(\(t<t'\)) Green's function. Using the Green's functions, we may
extend the integration limits to \(\pm\infty\):
\begin{equation}
  \tag{\ref{eqn:partitioned-tdse}**}
  \begin{aligned}
    \ket{\Psi_\pm(a)}
    &\equiv
      \propU_\pm(a,b)
      \ket{\Psi_\pm(b)} \\
    &\hphantom{=}-
      \im
      \int\diff{t}
      \greensfunction_\pm^+(a,t)
      \bloch_\mp
      \greensfunction_\mp^+(t,b)
      \ket{\Psi_\mp(b)} \\
    &=
      \propU_\pm(a,b)
      \ket{\Psi_\pm(b)} \\
    &\hphantom{=}-
      \im
      \int\diff{t}
      \greensfunction_\pm^+(a,t)
      \bloch_\mp
      \ket{\Psi(t)}
      \heaviside(t-b)
  \end{aligned}
\end{equation}
where we have additionally used that by
\eqref{eqn:subspace-hamiltonians-with-bloch},
\(\bloch_\pm\greensfunction_\pm^+(t,b)\ket{\Psi_\pm(b)} \equiv
\bloch_\pm\ket{\Psi(t)}\heaviside(t-b)\). Now, we compute the
fragmentation amplitude, in analogy with
\eqref{eqn:fragmentation-channels-equation-system}:
\begin{equation}
  \label{eqn:analytical-R-matrix-amplitude}
  a_{I\vec{v}}(T) =
  -\im\int\diff{t}
  \matrixel{I\vec{v}(T)}{\greensfunction_-^+(T,t)\bloch_+}{\Psi(t)},
\end{equation}
assuming no initial overlap, \(\braket{I\vec{v}}{\Psi_-(t_i)}=0\), as
well as complete localization of the wavefunction
\(\braket{\vec{r}}{\Psi(t)}\), for \(t\le t_i\). This formulation is
the one employed in analytical \(R\)-matrix theory
\cite{Torlina2012PRA,Torlina2012}, wherein the Green's function
\(\greensfunction_-^+(T,t)\) is subsequently approximated using e.g.\
eikonal-Volkov states \cite{Smirnova2008}. Here, we instead use
equations~\eqref{eqn:ionic-state-time-evolution},
\eqref{eqn:fragment-hamiltonian-matrix-elements}, and
\eqref{eqn:ion-channel-coupling-matrix} to write
\begin{equation}
  \label{eqn:greens-function-approximation}
  \greensfunction_-^+(T,t) \approx
  \heaviside(T-t)
  \ketbra{I\vec{v}(T)}{I'\vec{v}(t)}.
\end{equation}

In spherical symmetry, where \(\proj_+=1-\heaviside(r-a)\), and
employing the length gauge, the Bloch operator
\eqref{eqn:bloch-operator} is given by
\cite{Burke2005,Torlina2012PRA,Torlina2012}
\begin{equation}
  \label{eqn:bloch-operator-spherical-symmetry}
  \bloch_\pm
  \equiv
  \pm\frac{1}{2m_j}
  \delta(r_j-a)
  \left(
    \partial_{r_j} -
    \frac{b_0-1}{r_j}
  \right),
\end{equation}
where \(r_j\) and \(m_j\) are the radial coordinate and the mass of
the \(j\)th particle, respectively, \(a\) is the matching radius, and
\(b_0\) an arbitrary constant [which can be used to enforce the
desired boundary conditions of solutions to
\eqref{eqn:partitioned-tise} at \(r=a\)]. The Bloch operator in
spherical symmetry \eqref{eqn:bloch-operator-spherical-symmetry}
should be compared with \eqref{eqn:surface-term-current} and
\eqref{eqn:surface-integral-spherical-symmetry}. Figure~4 of
\cite{Burke2005} illustrates the special case of our
figure~\ref{fig:local-coordinate-system} to ionization and
disassociation of a molecule.

\section{Extension to Relativistic Quantum Mechanics}
\label{app:relativistic-extension}

The extension to relativistic quantum mechanics is straightforward,
provided that we make some simplifying assumptions:
\begin{enumerate}
\item Relativity is mostly important in the interior domain
  \(\domain\), affecting the energy structure of the bound states, due
  to the high speed of the electrons in the vicinity of the
  nuclei. The photofragments in typical experiments do not travel at
  any appreciable fraction of the speed of light, and hence relativity
  plays a minor rôle in the exterior domain \(\complementdomain\).
\item In the interior domain, the energy structure can be described by
  a Hamiltonian in the Dirac–Coulomb family of methods, i.e.\ with the
  Dirac Hamiltonian \(\Dirachamiltonian\) describing the motion of a
  single electron, and the nuclear--electron attraction, and
  electron--electron or nucleus--nucleus repulsion by the classical
  Coulomb potential. Corrections such as the Breit interaction and the
  Lamb shift may also be included. Importantly, we neglect pair
  production (the \emph{no pair} approximation).
\item In the exterior domain, the fragments obey the minimal coupling
  Dirac Hamiltonian \(\Dirachamiltonian(t)\).
\end{enumerate}
With these assumptions in place, ignoring the breaking of Lorentz
covariance of different reference frames by the introduction of the
partitioning of space into \(\domain\) and \(\complementdomain\)
becomes a negligible problem. We illustrate the procedure for single
ionization of one electron, the extension to the many-body problem is
completely analogous to the non-relativistic case.

The Dirac equation for one particle reads
\begin{equation}
  \label{eqn:dirac-equation}
  [\im\gamma^\mu\partial_\mu - mc]\psi = 0,
\end{equation}
where the \(\gamma\) matrices obey the Clifford algebra
\begin{equation}
  \anticomm{\gamma^\mu}{\gamma^\nu} =
  2\eta^{\mu\nu}\identity_4,
\end{equation}
and \(\eta^{\mu\nu}=\operatorname{diag}\bmat{+1 & -1 & -1 & -1}\) is
the Minkowski metric. The adjoint of the Dirac equation
\eqref{eqn:dirac-equation} is given by
\begin{equation}
  \label{eqn:adjoint-dirac-equation}
  \diracadjoint{\psi}
  [\im\overleftarrow{\partial_\mu}\gamma^\mu + mc] = 0,
\end{equation}
where \(\overleftarrow{\partial_\mu}\) indicates that the derivative
acts to the left, and the
\begin{equation}
  \diracadjoint{\psi} \defd
  \adjoint{\psi}
  \gamma^0
\end{equation}
denotes the Dirac adjoint.

The 4-current is defined as
\begin{equation}
  \label{eqn:dirac-current}
  j^\mu \defd
  \diracadjoint{\psi}
  \gamma^\mu
  \psi,
\end{equation}
and its divergence is found as
\begin{equation}
  \begin{aligned}
    \partial_\mu
    j^\mu
    &=
      \markterm{1}{(\partial_\mu
      \diracadjoint{\psi})}
      \gamma^\mu
      \psi +
      \diracadjoint{\psi}
      \gamma^\mu
      \markterm{2}{(\partial_\mu\psi)} \\
    &=
      \markterm{1}{(
      \im m c
      \diracadjoint{\psi}
      )}
      \psi -
      \diracadjoint{\psi}
      \markterm{2}{(\im m c \psi)} = 0,
  \end{aligned}
\end{equation}
where in the second step we have inserted the Dirac equation
\eqref{eqn:dirac-equation} and its adjoint
\eqref{eqn:adjoint-dirac-equation}, and found through the cancellation
that the divergence of the 4-current vanishes:
\begin{equation*}
  \partial_\mu j^\mu = 0,
\end{equation*}
which we can rewrite in 3-vector notation as
\begin{equation}
  \label{eqn:dirac-continuity-equation}
  \partial_t
  \adjoint{\psi}\psi +
  \partial_{\vec{r}}\cdot\vec{j} = 0,
\end{equation}
since the \emph{probability density} \(j^0\) and the probability
current \(\vec{j}\) are given by
\begin{equation}
  \begin{aligned}
    j^0
    &=
      \diracadjoint{\psi}
      \gamma^0
      \psi =
      \adjoint{\psi}\psi,
    &
      \vec{j}
    &=
      \diracadjoint{\psi}
      \vec{\gamma}
      \psi.
  \end{aligned}
\end{equation}
Now, in full analogy with the non-relativistic case, we may rewrite
\eqref{eqn:dirac-continuity-equation} on integral form, employ the
divergence theorem, and perform the time integral to find the
probability in the exterior domain \(\complementdomain\) as
\begin{equation}
  \label{eqn:dirac-continuity-equation-integral-form}
  \tag{\ref{eqn:dirac-continuity-equation}*}
  \int_{\complementdomain}\diff{\vec{r}}
  \adjoint{\psi}
  \psi =
  \int\diff{t}
  \int_{\domain}\diff{\vec{r}}
  \partial_{\vec{r}}\cdot\vec{j} =
  \int\diff{t}
  \oint_{\domainboundary}
  \vec{j}
  \cdot
  \diff{\vec{S}},
\end{equation}
which is exactly analogous to the non-relativistic case
\eqref{eqn:probability}.

The introduction of an external electromagnetic field amounts to
replacing the derivative in the Dirac equation
\eqref{eqn:dirac-equation} and its adjoint
\eqref{eqn:adjoint-dirac-equation} with the covariant derivative in
minimal coupling according to
\begin{equation}
  \partial_\mu \to
  D_\mu \defd
  \partial_\mu +
  \im q A_\mu,
\end{equation}
where the 4-potential is given by
\begin{equation}
  A = \bmat{\phi & -\vec{A}}.
\end{equation}
The continuity equation \eqref{eqn:dirac-continuity-equation} still
holds as-is [as does
\eqref{eqn:dirac-continuity-equation-integral-form}]. The vector
potential \(\vec{A}\) does not appear explicitly in the Dirac current
\eqref{eqn:dirac-current}, but in the non-relativistic limit, upon
elimination of the small component via the Dirac equation
\eqref{eqn:dirac-equation}, the non-relativistic probability current
\eqref{eqn:current} gains an extra term.


\bibliographystyle{apsrev4-2}
\bibliography{\bibliographyfile}

\end{document}